\documentclass[12pt]{iopart}
\usepackage{graphicx}
\usepackage{amssymb}

\usepackage{epstopdf}
\usepackage[square,sort,comma,numbers]{natbib}
\begin{document}
\def\aligned{\vcenter\bgroup\let\\\cr
\halign\bgroup&\hfil${}##{}$&${}##{}$\hfil\cr}
\def\endaligned{\crcr\egroup\egroup}
\title{Engineering passive swimmers by shaking liquids
}
\author{M Laumann$^1$, A F\"ortsch$^1$, E Kanso$^2$ and W Zimmermann$^1$}
\address{Theoretische Physik I, Universit\"at Bayreuth, Bayreuth, Germany$^1$\\
 Aerospace and Mechanical Engineering, University of Southern California, Los Angeles, California, USA$^2$}
\ead{walter.zimmermann@uni-bayreuth.de}

\begin{abstract} The locomotion and design of  microswimmers are topical issues of current fundamental and applied research.  
In addition to numerous living and artificial active microswimmers, a passive microswimmer  was identified only recently:
a soft, $\Lambda$-shaped, non-buoyant particle propagates in a shaken liquid of zero-mean  
velocity [Jo {\it et al.} Phys. Rev. E {\bf 94}, 063116 (2016)]. We show that this novel
passive locomotion mechanism  works  for realistic non-buoyant, asymmetric  Janus microcapsules as well.
According to our analytical approximation, this
locomotion requires a symmetry breaking  caused by different Stokes drags of  soft particles during the two
half periods of the oscillatory liquid motion. 
It is the intrinsic anisotropy of Janus capsules and $\Lambda$-shaped particles
that break this symmetry for sinusoidal liquid motion. Further, we show   
that this  passive locomotion mechanism also works  for the wider class of
symmetric soft particles, e.g., capsules, 
by breaking the symmetry via an appropriate liquid shaking.
The swimming direction can be uniquely selected by a
suitable  choice of the liquid motion. 
Numerical studies, including lattice Boltzmann simulations, also show  that 
this locomotion can outweigh gravity, i.e.,
non-buoyant particles may be either elevated in  shaken liquids 
or concentrated at the bottom of a container.
This novel propulsion mechanism is relevant to many applications, 
including the sorting of soft particles 
like healthy and malignant (cancer) cells, which serves medical purposes, or the use of non-buoyant
soft particles as directed microswimmers .
\end{abstract}

%
%
\submitto{\NJP}
%
\maketitle
%
%

\section{Introduction}

Biological microswimmers and their artificial  counterparts attract a great deal of attention
in research
both for their fundamental relevance and their potential applications in a variety of  physical, 
biological, chemical or biomedical applications (see e.g., \cite{Goldstein:2012.2,Lauga:2009.1,Guasto:2012.1,Bechinger:2016.1,Lauga:2016.1}).
Several studies focus on the dynamics of soft particles in microflows, such as capsules and red blood cells
\cite{Quake:2005.1,Beebe:2014.1,Kumar_S:2015.1,DiCarlo:2014.1,Secomb:2017.1}.
Their exploration and  understanding inspires, among others,  passive microswimmers that are indirectly driven by a
time-dependent liquid motion. 
An example is a recently identified  inertia-driven, passive microswimmer:
A non-buoyant asymmetric soft microparticle in oscillatory liquid motion of zero mean displacement was studied in 
\cite{Kanso:2016.1,Ishikawa:2018.1}. Here we show
how this  inertia-driven locomotion mechanism can be generalized to the much wider class of homogeneous, soft particles, such as capsules,
by engineering an appropriate  time-dependent liquid motion.

Mechanisms that underly the propulsion of microswimmers include the propulsion
 via  chemical reactions on the anisotropic surface of Janus particles, by magnetic fields or acoustic fields
(see e.g., \cite{Bechinger:2016.1}). Common propulsion mechanisms of 
microorganisms at low Reynolds number are  periodic motions of flagella, cilia or 
the deformation of the body shape (amoeboid motion)
\cite{Lauga:2009.1,Guasto:2012.1,GoldsteinR:2015.1,Lauga:2016.1,Wu:2015.1,Wu:2016.1,Farutin:2013.1}. 
To achieve a net displacement at these length scales
the mechanism has to be non-reciprocal to break Purcell's scallop theorem \cite{Purcell:1977.1,Lauga:2009.1}.

The non-reciprocal motion of biological swimmers inspired also passive artificial microswimmers recently.
One example is a  soft Janus capsule in a temporally periodic linear shear flow at low Reynolds number,
whereby the intrinsically  asymmetric Janus particle is   propelled  perpendicular to the streamlines \cite{Laumann:2017.1}.
This type of passive swimming
and the theoretical model of a brake controlled triangle \cite{Olla:2010.1}  
are similar to cross-stream migration of droplets and soft particles in 
stationary low Reynolds number Poiseuille flows \cite{Leal:1980.1,Chakraborty:2015.1,Kaoui:2008.1,Misbah:2008.1,Bagchi:2008.1}.
Other recent  studies identified the  finite inertia of soft particles in 
oscillatory homogeneous liquid motion  as a crucial property for passive swimming \cite{Kanso:2016.1,Ishikawa:2018.1}. 
The non-reciprocal  body shape and
therefore the different Stokes drag in both half periods of the periodic liquid motion is the driving force of 
these novel locomotion mechanism.
The first inertia driven particle locomotion at low Reynolds number  was demonstrated for
a soft, asymmetric, $\Lambda$-shaped particle  
in a shaken liquid  \cite{Kanso:2016.1}.
This  was extended to an internally structured capsule  with an inhomogeneous
mass distribution in a gravitation field  \cite{Ishikawa:2018.1}. 
\begin{figure*}[htb]
\begin{center}
 \includegraphics[width=\textwidth]{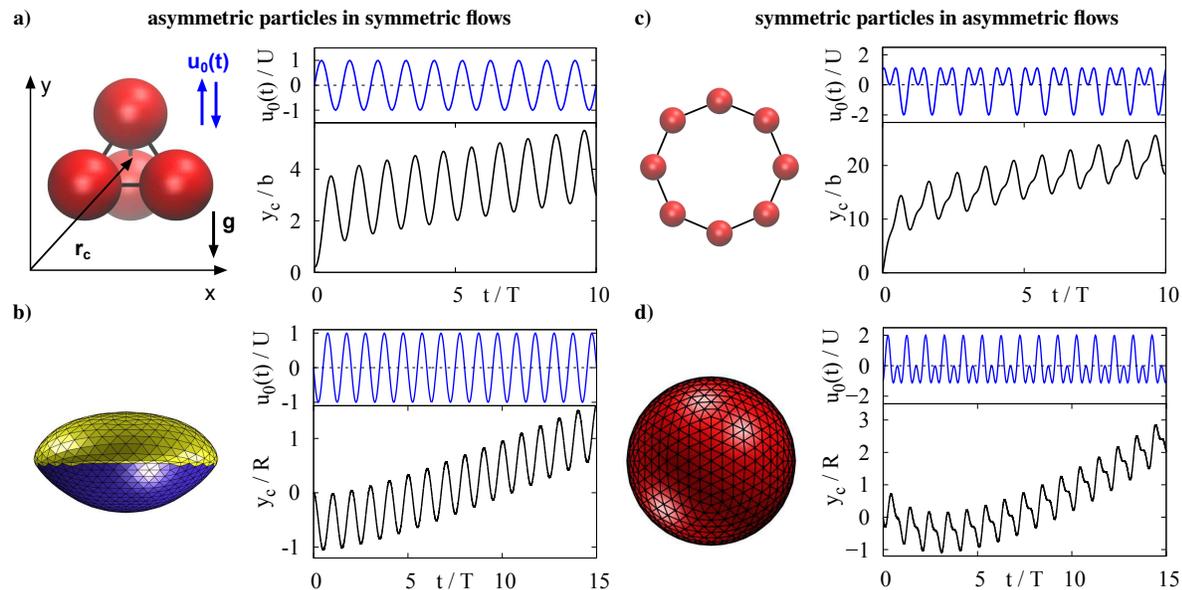}
\end{center}
\vspace{-0.7cm}
\caption{Sketch of four different particles with a mass density different from that of the  surrounding liquid.
The liquid   shaken with the velocity
${\bf u}_0(t)=u_0(t){\bf e}_y$ causes an  inertia-driven particle locomotion.
The actuation is indicated by  the motion of the particle's center of mass ${\bf r}_{c}(t)=y_c(t) {\bf e}_y$.
Part a) shows an asymmetric bead-spring tetrahedron and part b) an asymmetric Janus capsule with  different 
stiffness of each half  (soft part yellow). Both  asymmetric particles 
are considered in a sinusoidal velocity of the liquid with period $T$,
as indicated by the upper blue curves in a) and b).
We observe for both particles a net motion along the direction of shaking  against gravitation, as described by the black curves in a) and b). 
The ring in c) and the symmetric capsule in d) are shaken by  non-symmetric velocities as indicated by the blue curves in 
part c) and d). Also the ring and the symmetric capsule 
show a net progress  against gravity  as indicated by
the black curves in c) and d).  Parameters are given in sections \ref{models} and \ref{results}.
}
\label{Sketchsink}
\end{figure*}

 In this work we show that
 the inertia induced 
 passive swimming  of realistic and experimentally available soft particles in oscillatory liquid motion 
  can  outweigh gravitation. We show this at first for
an  Janus capsule with an asymmetric elasticity (see e.g. \cite{Chen:2015.1}). 
We explain
that an intrinsic particle asymmetry is not required for passive swimming and  
we demonstrate that the inertia driven particle propulsion works also  for the much wider class of 
 homogeneous and symmetric soft particles, such as soft capsules. 
 This is achieved by appropriately engineering the time-dependence of shaking  the liquid.
 The time-dependence of the shaking  determines also
 the direction of passive swimming.  This motion  is distinct from particle 
 locomotion in oscillatory flows at finite Re, where propulsion is 
 related to streaming flows and a fluid jet in the wake of the swimmer \cite{SwiftMR:2015.1}.

The work is organized as follows: In section \ref{models} we describe the modeling and simulation of the particles sketched in 
figure \ref{Sketchsink}. We show in section \ref{sec_explanation} by an approximate analytical approach, that 
the locomotion of non-buoyant soft particles in a periodically oscillating fluid motion requires the symmetry breaking 
caused by different  particle deformations and Stokes drags
during the two half-periods of the shaking of the liquid.
The analytical results are confirmed in section \ref{results}
by numerical simulations of the bead-spring models 
and capsules shown in figure \ref{Sketchsink}.
We study an asymmetric bead-spring tetrahedron in a  sinusoidal liquid motion and a symmetric semiflexible bead-spring ring 
in a non-symmetric periodic liquid motion for a wide  parameter range. The results of these simulations  are complemented and verified
by  Lattice Boltzmann simulations of  realistic soft asymmetric Janus capsules and symmetric capsules. For instance, we
provide parameter ranges where the passive locomotion mechanism outweighs gravitation.
Discussions of the results and the conclusions are given in sec. \ref{conclusion}.

\section{Model and Approach}\label{models}
The dynamics of four  deformable particles  in a shaken liquid is investigated by taking into account particle inertia.
We use two asymmetric particles, namely a
bead-spring tetrahedron composed of four beads, and a Janus capsule, as sketched in figures~\ref{Sketchsink} a) and  b),
respectively. As examples of common symmetric
particles we choose a bead-spring ring, as shown in  figure \ref{Sketchsink}c), and 
a symmetric capsule, as shown in figure \ref{Sketchsink}d). 
The positions and the motion of the
beads of the ring are restricted to the $(x,y)$ plane.

The  shaking velocity of the liquid is given by
\begin{eqnarray}
 \mathbf{u}_0(t)&=u_0(t)\hat{e}_y =U\, \left[\sin\left(\omega t\right)+\varepsilon\cos\left(2 \omega t\right) \right]\hat{e}_y\,,
   \label{eq_flow}
\end{eqnarray}
with the frequency $\omega=2\pi/T$ and a vanishing mean velocity $\langle  \mathbf{u}_0(t) \rangle =0$.
For $\varepsilon=0$ the velocity is  sinusoidal and antisymmetric with respect to a shift $t \to t +T/2$, i.e.,
 ${\bf u}_0(t)=-{\bf u}_0(t+T/2)$. 
 For $\varepsilon \not =0$ this symmetry is broken and the  velocity of the liquid is non-symmetric 
as indicated by the blue curves in figures~\ref{Sketchsink}c) and \ref{Sketchsink}d).

In section \ref{springmod} we describe the modeling of the bead-spring models and the capsules. 
In section \ref{models_oseen}, we present the equations of motion of the bead spring models,
the Maxey and Riley equations \cite{MaxeyRiley:1983} for several  beads. They take the particle inertia
into account  and are extended by the hydrodynamic
particle-particle interaction via the dynamical Oseen-tensor.
The Lattice-Boltzmann-Method (LBM)  for the particle simulations is explained 
in section \ref{models_LBM}.

\subsection{Modeling the bead-spring models and the capsules }\label{springmod}
The beads of the bead-spring models have the mass $m_i$. Their mass density $\rho_i$ may be different 
from the mass density of the fluid, $\rho_f \not = \rho_i$.
With the gravitational force  along the negative $y$ direction, this leads 
to the buoyancy force 
\begin{equation}
{\bf F}_{g,i}=-F_{g,i}{\bf e}_y, 
\end{equation}
which acts on  a particle immersed in the liquid with 
\begin{equation}
 F_{g,i}=gV_i(\rho_i-\rho_f)=g(m_i-m_f)\,.
\end{equation}

The tetrahedron in figure \ref{Sketchsink}a) 
consists of $N=4$ beads at  positions $\mathbf{r}_i$. The beads have the same
radius $a$, but may have different masses. They are connected
by springs with the stiffness $k$. The center of mass is given by
\begin{equation}
\mathbf{r}_c=\frac{1}{\sum_i m_i}\sum_i m_i\mathbf{r}_i\,.
\end{equation} 
Each bead experiences a force that is composed of the
buoyancy force ${\bf F}_g$ and forces imposed by  springs,
\begin{equation}
\label{eq_fp}
\mathbf{F}_i^{(P)}=-F_{g,i}\hat{e}_y-\nabla_i V_{spring} 
\end{equation}
with the spring potential
\begin{equation}
V_{spring}=\sum\limits_{i,j\neq i}^{N}k(|\mathbf{r}_i-\mathbf{r}_{j}|-b)^2\,\\
\label{eq_m}
\end{equation}
and the undistorted spring length $b$.

Also for the bead-spring model shown in figure \ref{Sketchsink}c) (with $N=8$ beads)
the neighboring beads are connected by Hookean springs. In addition to equation (\ref{eq_fp})
a  bending potential with the stiffness $\kappa$ is taken into account
 \begin{eqnarray}
 \label{eq_fp_bend}
 V_{bend}=-\frac{ \kappa} {2} \sum \limits_{i=1}^N \ln\left( 1 + \cos \beta_i \right)\,
 \end{eqnarray}
 where ${\bf R}_i={\bf r}_{i} - {\bf r}_{i+1}$ is the the bond vector between the next-neighbor 
 beads $i$ and $i+1$ and the  angle $\beta_i$ is defined 
 via $\cos \beta_i  = {\bf{e}}_{R_{i-1}} \cdot {\bf{e}}_{R_{i}}$ with the bond unit 
 vectors  ${\bf{e}}_{R_{j}} = {\bf R}_j/R_j$. This bending potential 
 causes a circular ring shape in equilibrium.

The capsules are modeled by discretizing their surface with $N=642$, which is done iteratively as described in more detail in
Ref.~\cite{Krueger:2011.1}.
We assume that the surface is thin and has a constant surface shear elastic modulus $G_s$. 
In this case the relation between the deformation and the forces is given by the neo-Hookean 
law described by the potential $V_{NH}$ (for details we refer to ~\cite{ramanujan:1998.1,BarthesBiesel:2016.1}). 
Furthermore a bending potential $V_{b}$ is assumed \cite{Gompper:1996.1,Krueger:2013.1}, which is given by 
\begin{eqnarray}\label{eq_cap_Vb}
V_{b} = \frac{\kappa_c}{ 2} \sum \limits_{i,j}  \beta_{i,j}^2\,,
\end{eqnarray} 
where $\kappa_c$ denotes the bending stiffness and $\beta_{i,j}$ is the angle between the normal vectors 
of neighboring triangles. 

For  Janus capsules the  stiffness is different in both halves of the capsule,
as indicated  in figure. \ref{Sketchsink}(c).
We use a penalty force to keep the capsule's volume ${\cal V}(t)$ 
close to the reference volume ${\cal V}_0$ during the simulations. Its potential $V_v$ is given by
\begin{eqnarray}\label{eq_cap_V_Vol}
V_{v} = \frac{k_v}{{\cal V}_0} ({\cal V}(t) - {\cal V}_0)^2\,
\end{eqnarray} 
with the rigidity $k_v$ \cite{Krueger:2013.1}. The complete potential related to  the forces acting on the capsule is given by
 \begin{eqnarray}
 \label{eq_pot_caps}
 V({\bf r}) = V_{NH} + V_{b} + V_{v}.
 \end{eqnarray}

\subsection{Maxey and Riley equations, including the dynamic Oseen-Tensor}\label{models_oseen}
The dynamics of the beads 
is described by the equations for the particle velocities ${\bf v}_i$ of Maxey and Riley \cite{MaxeyRiley:1983}. The flow in presence of the particles is denoted by ${\bf u}({\bf r})$ and the flow field without the particles as ${\bf u}_0({\bf r})$ (given by equation (\ref{eq_flow})). The flow in presence of the particles must fulfill the conditions
\begin{eqnarray}
\rho_f\frac{\partial \bf u}{\partial t}&=&\rho{\bf g}-\nabla p+\eta\Delta {\bf u},\\
\nabla\cdot {\bf u}&=&0\\
 {\bf u}({\bf r})&=&{\bf u}_0({\bf r}) \mbox{ at } |{\bf r}|\rightarrow \infty
\end{eqnarray}
and no slip boundary conditions on the surface of the beads. The equation of motion of the spheres is
\begin{equation}
m_i\frac{d {\bf v}_i}{dt}=m_p{\bf g}+\oint\sigma d{\bf A}_i\,, \label{eq_m_bead_1}
\end{equation}
i.e. the integral over the stress tensor must be calculated. In this equation \ref{eq_m_bead_1}, the inertia of the beads is considered by the term $m_i\frac{d {\bf v}_i}{dt}$. This inertia of the particle is often neglected in bead spring models, but becomes important e.g. if the external flow has a high enough frequency (see sec. \ref{eq_m_bead_1}). This inertia can also occur if the Reynolds number of the flow around the particle is zero. The Reynolds number influences only the flow and thus the integral over the stress tensor in eq. (\ref{eq_m_bead_1}). To calculate the integral in equation \ref{eq_m_bead_1}, a small Reynolds number, defined by $\mbox{Re}=\rho_f aw/\eta$ with the relative velocity ${\bf w}$ between the fluid velocity ${\bf u}_0$ and the particles velocity ${\bf v}$ is assumed, which allows to neglect the advective terms in the Navier-Stokes equation. We keep the time-derivative ${\partial_t \bf u}$ due to the high Strouhal number. The calculation of the integral over the stress tensor is given in \cite{MaxeyRiley:1983} and leads to the following forces on the beads:

A bead  experiences besides ${\bf F}_i^{(P)}$  the inertial force
\begin{equation}
 \label{eq_fd0}
\mathbf{F}_i^{(0)}=m_{f,i}\frac{d \mathbf{u}_i}{dt}\,
\end{equation}
caused by the liquid acceleration at the position ${\bf r}_i$ of the bead. Note that the liquid velocity includes
the externally imposed homogeneous liquid motion ${\bf u}_0(t)$ described in equation (\ref{eq_flow}) and the flow perturbations caused by the motion of all other particles
with respect to the liquid.
Furthermore, the force ${\bf F}_i^{(1)}$ created by the difference between the particle velocity ${\bf v}_i$ and the liquid velocity ${\bf u}_i$
must be considered.
This is composed of
three contributions, the added mass, 
the Stokes drag and the Basset force,
\begin{eqnarray}
\label{eq_fd1}
\mathbf{F}_i^{(1)}&=&-\frac{1}{2}m_{f,i}\frac{d}{dt}\left(\mathbf{v}_i-\mathbf{u}_i\right)-\zeta_b(\mathbf{v}_i-\mathbf{u}_i)\nonumber\\
&&\qquad -6\pi\eta a^2 \int_0^t d\tau \frac{\frac{d}{d\tau}\left[ \mathbf{v}_i(\tau)-\mathbf{u}_i(\tau)\right]}{\sqrt{\pi\nu(t-\tau)}}\,,
 \end{eqnarray}
 with the Stokes drag coefficient $\zeta_b=6\pi\eta a$. 
 Alltogether we obtain the dynamical equation for the velocity of the $i$th bead
 \begin{eqnarray}
  \label{eq_newton}
 m_i\frac{d\mathbf{v}_i}{dt}= \mathbf{F}_i^{(0)}&+&\mathbf{F}_i^{(1)}+\mathbf{F}_i^{(P)}\,.
 \end{eqnarray}
The flow disturbances at ${\bf r}_i$ caused by all the other beads 
are determined via the dynamic Oseen tensor \cite{Ignacio:1995}, which is 
the Greens function of the time-dependent linear Stokes equation. 
This provides the flow at the $i$th bead
 \begin{equation}
\mathbf{u}_i=\mathbf{u}_0(t)+\sum\limits_{j\neq i}\int\limits_0^t dt'\ \mathbf{H}_{i,j}(t')\cdot\mathbf{F}_j^{(1)}(t')\,.\\
\label{eq_ui}
 \end{equation}
 For the explicit expression of $\mathbf{H}_{i,j}(t')$
we refer to  \ref{models_oseen_SI}.

Equation (\ref{eq_fd1}) is solved  numerically for the bead-spring tetrahedron as shown in figure \ref{Sketchsink}a)
and for the bead-spring ring shown in figure \ref{Sketchsink}c) 
by using  a Runge-Kutta-scheme of fourth order.
The dimensionless parameters given below are used for simulations of equation (\ref{eq_newton}) for  the bead-spring tetrahedron 
and the bead-spring ring. 
   These parameters  can be converted to SI units if the dimensionless time is multiplied by the factor $s_t=1\ \mbox{ms}$, 
   the length by $s_l=50\ \mu \mbox{m}$ and mass  by $s_{m}=5.2\cdot 10^{-13}\ \mbox{kg}$. 
This leads to the density and viscosity of water ($\rho_{water}=1000\ \mbox{kg}/\mbox{m}^3$, $\eta_{water}=1 \mbox{  mPas }$) and the correct gravitational acceleration $g\approx 10\ \mbox{m}/\mbox{s}^2$.

The {Parameters} used in simulations of the bead-spring tetrahedron are:
number of beads $N=4$, 
bead radius $a=0.1$, equilibrium spring length $b=0.25$, spring stiffness $k=15000$, 
mass density $\rho_i=3600$ of a bead, mass density of the fluid $\rho_f = 240$, fluid viscosity $\eta=100.0$, 
amplitude of the shaking velocity $U = 10.0$ in equation (\ref{eq_flow}), 
asymmetry parameter $\varepsilon=0$,
shaking period {$T=0.4$}, 
gravitational acceleration $g= 0.21$ and
time step  $dt=2.5\cdot10^{-4}$ in numerical integrations of equation (\ref{eq_newton}). 

The {Parameters} used in simulations of the semiflexible bead-spring ring are:
number of beads $N=8$, 
bead radius $a=0.1$, 
equilibrium spring length $b=0.5$, 
spring stiffness $k=2000$, 
bending stiffness $\kappa=500$, 
mass density $\rho_i=3600$ of a bead,
mass density of the liquid $\rho_f = 240$, 
viscosity of the liquid $\eta=100$, 
amplitude  $U = 20$  of the shaking velocity in equation (\ref{eq_flow}),
asymmetry parameter $\varepsilon=0.8$, 
shaking period  $T=0.4$, 
gravitational acceleration $g= 0.21$ and
time step  $dt=2.5\cdot10^{-4}$ in numerical integrations of equation (\ref{eq_newton}).

Especially the flow amplitude is varied which leads to Reynolds numbers $\mbox{Re}=\rho_f U b /\eta$ ranging from $0$ to $6$ and Strouhal numbers $\mbox{St}=b/(TU)$ ranging from $0.06$ to $\infty$. Due to the high range of the Strouhal number the time-derivative in the Navier-Stokes equation is kept. We neglect the advective terms in the Navier-Stokes equation to show that the locomotion of the particles is not induced by the non-linearity of the Navier-Stokes equation but by the simple Stokes drag. To verify this approximation the qualitative results are compared with simulations of the lattice-Boltzmann Method, which solves the full Navier-Stokes equation.

\subsection{The Lattice Boltzmann Method}\label{models_LBM}

We use also the LBM to simulate the full Navier–Stokes equation including the dynamics of the particles. 
We utilize the common D3Q19 lattice-Boltzmann method (LBM) to 
simulate the distribution $f(\mathbf{x},t)$  of the fluid elements
on a 3D grid of positions $\mathbf{x}_i=(x,y,z)$ along the discrete 
directions  $\mathbf{c}_i$ ($i=0,\dots,19$) \cite{Aidun:2010}.
The lattice constants are $\Delta x=1$ for spatial 
and  $\Delta t=1$ for temporal discretization. 
The evolution of the distribution function 
is governed by the discrete Boltzmann equation
\begin{equation}
\label{Boltz1}
    f_i(\mathbf{x}+\mathbf{c}_i\Delta t,t+\Delta t) = f_i(\mathbf{x},t) + \mathcal{C},
\end{equation}
where $\mathcal{C}$ defines the collision operator. Walls are incorporated by the standard bounce back scheme (bb) \cite{Aidun:1998,dHumieres:2002} by
adding
the contribution $\mathcal{W}=2w_i\rho \frac{\mathbf{c}_i \cdot \mathbf{u}_w}{c_s^2}$ for wall links to equation (\ref{Boltz1}) \cite{Ladd:1994:1,dHumieres:2002}, 
where $\mathbf{u}_w$ is the wall velocity. The weighting factor $w_i$ and the speed-of sound $c_s$
are constants for the chosen set of velocity directions \cite{Aidun:2010}.

\paragraph{Tetrahedron dynamics:} For the simulations of the tetrahedron, 
the Bhatnagar-Gross-Krook (BGK) collision operator
\begin{eqnarray}
    \mathcal{C}= -\frac{1}{\tau} &\left[ f_i(\mathbf{x},t) - f^{eq}_i(\mathbf{x},t) \right] + \mathcal{F}
\end{eqnarray}
 is extend by  the Guo force-coupling $\mathcal{F}=\Delta t \left( 1 - \frac{1}{2\tau} \right) w_i \left[ \frac{c_i -
\mathbf{u}}{c_s^2} + \frac{(\mathbf{c_i} \cdot \mathbf{u})}{c_s^4} \mathbf{c}_i \right] \cdot \mathbf{F}^{(e)}$ for external volume forces 
$\mathbf{F}^{(e)}$ \cite{Guo:2002}. $f^{eq}$ is an expansion of the Maxwell-Boltzmann distribution and $\tau$ is the relaxation parameter.
The macroscopic density and momentum are obtained 
from the first two moments via $\rho = \sum_i f_i$ and $\rho \mathbf{u} = \sum_i \mathbf{c}_i f_i + \frac{\Delta t}{2} \mathbf{F}^{(e)}$,
respectively. The viscosity of the fluid is given by $\nu = c_s^2 \Delta t \left(\tau - 1/2 \right)$. 
The hard spheres are implemented as moving walls according to \cite{Aidun:1998}, 
with an additional lubrication-correction for squeezing motion of near particles, as discussed in \cite{Ladd:2001}. 
This simulations are used to compare the Oseen simulations and the LBM simulations (see also \ref{tetra_SI}).

\paragraph{Capsule dynamics:} For the simulations of capsules, an adapted LBM-scheme of the multi-relaxation 
time LBM for a spatially dependent density is used \cite{Shao:2015}. The time evolution of the 
mean density $\rho_0(\mathbf{x},t)=\sum_i f_i + \frac{1}{2} \mathbf{u} \nabla \rho \Delta t$, 
the local density $\rho(\mathbf{x},t)$ and its gradient $\nabla \rho$ is used as input for the collision operator
\begin{eqnarray}
    \mathcal{C} = &-S_{il}\left[f_l(\mathbf{x},t) - f_l^{eq}(\mathbf{x},t) \right] \nonumber \\
    &+F_i(\mathbf{x},t) - \frac{1}{2} S_{il} F_l (\mathbf{x},t).
\end{eqnarray}
For the collision matrix $\mathbf{S}$ and its corresponding transformation matrix we use the set given in \cite{Premnath:2007}. The correction term $F_i(\mathbf{x},t)=\Delta t \frac{(\mathbf{c}_i - \mathbf{u})}{c_s^2} \cdot \left[ \nabla \rho c_s^2 (\Gamma_i-w_i) + \mathbf{F}^{(e)} \Gamma_i \right]$ 
accounts for the density inhomogeneity and external forces, 
with $\Gamma_i=w_i \left[1+\frac{\mathbf{e}_i\cdot \mathbf{u}}{c_s^2} + \frac{(\mathbf{c}_i\cdot \mathbf{u})^2}{2 c_s^4} - \frac{|\mathbf{u}|^2}{2c_s^2}\right]$ \cite{Shao:2015}. 
The fluid velocity is linked to the density $\rho$ via the second moment $\rho \mathbf{u} = \sum_i f_i \mathbf{c}_i + \frac{1}{2} \mathbf{F}^{(e)}\Delta t$. The equilibrium distribution has the form $f_l^{eq}(\mathbf{x},t)=w_l\left[\rho_0 + \rho\left( \frac{(\mathbf{c}_l \cdot u)}{c_s^2} + \frac{(\mathbf{c}_l\cdot \mathbf{u})^2}{2 c_s^4} - \frac{|\mathbf{u}|^2}{2 c_s^2} \right) \right]$. The capsule mesh is coupled to the LBM-grid via the immersed-boundary method using the four-point stencil \cite{Peskin:2002}. 
The calculation of the field for the density $\rho(\mathbf{x},t)$ used in \cite{Shao:2015} is replaced by tracking nodes inside the capsule and setting $\rho(\mathbf{x},t)$ as $\rho_{capsule}$ inside and $\rho_{fluid}$ outside of the membrane and updating the capsule surface via the membrane-forces.

\paragraph{Oscillating flow.}
To drive the oscillating flow, an external (volume-)force 
$\mathbf{F}^{(e)}_{flow} = U  \rho\,\omega [\cos(t \omega) - 2 \varepsilon  \sin( 2  t \omega) ] \hat{\mathbf{e}}_y$ is applied to the LBM. 
To screen hydrodynamic self-interaction, we use bb walls in $x$ and $z$-direction with velocity $\mathbf{u}_w(t) = \mathbf{u}(t)$ 
to ensure Dirichlet boundary conditions of the flow.

\paragraph{Parameters and unit-conversion.}
The used LBM parameters can be obtained from the SI parameters via the conversion 
values for length $s_{{L}}=7.57\cdot 10^{-7}\mbox{m}$, mass 
$s_{{M}}=4.348\cdot 10^{-16} \mbox{kg}$ and time $s_{{T}}=4.54\cdot 10^{-8} \mbox{s}$. 
All LBM simulations are performed with a viscosity $\eta=\eta_{water}=1 \mbox{  mPas }$, gravitational acceleration 
$g=9.81 \mbox{m}/\mbox{s}^2$, fluid density $\rho = \rho_{water}=1000 \frac{\mbox{kg}}{\mbox{m}^3}$ and
$k_v = 2.78\cdot10^5\frac{\mbox{kg}} {\mbox{s}^2\mbox{m}}$. 
The amplitude of the liquid's velocity is $U=0.5\frac{\mbox{m}}{\mbox{s}}$ and the period is $T=90\mu\mbox{s}$
if not given otherwise. The cubic simulation box has a length of $1.14\times 10^{-4}$m.

\section{Inertia driven actuation: Approximate analytical results}
\label{sec_explanation}

Soft particles are periodically deformed in shaken liquids, which  causes a time-dependent viscous 
drag coefficient of the particle. How this deformability drives  passive swimming of a particle in a shaken liquid
is determined by an approximate analytical approach.

We discuss here a particle with a drag coefficient $\zeta_{tot}$.  This already simplifies
the dynamical equation  (\ref{eq_newton}). We further neglect the Basset force and the added mass in equation (\ref{eq_fd1}) but take the force $\mathbf{F}^{(0)}$
and the dominant viscous drag contribution to $\mathbf{F}^{(1)}$ into account. 
In this case we obtain the approximate dynamical equation for
the velocity of a stiff  particle 
\begin{equation} 
 M \frac{d \mathbf{v}(t)}{dt}=\zeta_{tot} \left[\mathbf{u}_0(t)-\mathbf{v}(t)\right] + M_f \frac{d \mathbf{u}_0(t)}{dt}\,,\label{max_riley_approx}
\end{equation}
 with the particle mass $M$, the displaced fluid mass $M_f$ and  the constant Stokes drag   coefficient  $\zeta_{tot}$.
 To justify the validity of this approximations we compare them with the full numerical results in the next section.
 
For a sinusoidal liquid motion ${\bf u}_0(t)$ as described by equation (\ref{eq_flow}) with  $\varepsilon=0$
the    solution of  equation (\ref{max_riley_approx})  is  $\mathbf{v}(t)=v(t)\hat{e}_y$ 
with
\begin{eqnarray}
\label{singlevt}
 & v &(t)  =  C e^{-\frac{\zeta_{tot}}{M} t}+A\sin\left(\omega t+\phi\right)\,,
 \end{eqnarray}
  whereby 
 \begin{eqnarray}
 A &  = & U \sqrt{\frac{M_f^2\omega^2+\zeta^2_{tot}}{M^2\omega^2 +\zeta^2_{tot}}}=U \sqrt{\frac{\left({M_f\omega}/{\zeta_{tot}}\right)^2+1}{\left({M\omega}/{\zeta_{tot}}\right)^2 +1}}
 \label{eqn_def_A}
 \end{eqnarray}
 is the amplitude of the particle oscillation 
 and the phase shift relative to the time-dependent liquid motion is given by
 \begin{eqnarray}
 \phi &  = & -\arctan\left(\frac{\zeta_{tot}\omega(M-M_f)}{MM_f\omega^2+\zeta^2_{tot}}\right).
 \label{eqn_def_phi}
\end{eqnarray}
 The exponential contribution to equation 
 (\ref{singlevt}) includes the relaxation time  $\tau_v=\frac{M}{\zeta_{tot}}$ that
 the particle needs to adjust its velocity ${\bf v}$ 
  to the velocity of the liquid ${\bf u}_0$.
 We approximate this time scale by
\begin{equation}
\tau_v=\frac{M}{\zeta_{tot}}\approx\frac{4m}{4\zeta_b}=\frac{m}{\zeta_b}\,
\label{def_tau_v}
\end{equation}
 for the bead spring models and for the capsule by
\begin{equation}
\label{def_tau_v_caps}
\tau_v=\frac{M}{\zeta_{tot}}=\frac{2 R^2\rho}{9\eta}
\end{equation}
 with
\begin{equation}
\zeta_{tot}=6\pi \eta R\,,\quad V=\frac{4}{3}\pi R^3\,,  \quad M=\rho V.
\end{equation}
We discuss in the following the case $M\geq M_f$ (but also $M<M_f$ is possible).
For a high friction or slow frequency, i.e.  $\frac{M}{\zeta_{tot}}\omega  \ll 1$,
the particle velocity adjusts rather quickly to the liquid motion. This means the particle quickly adapts to the motion of the liquid, i.e. $A\rightarrow U$ and $\phi\rightarrow0$ (cf. equations (\ref{eqn_def_A}) and (\ref{eqn_def_phi})) and the particle's inertia is negligible in this case.

\begin{figure*}[htb]
\begin{center}
\includegraphics[width=0.75\textwidth]{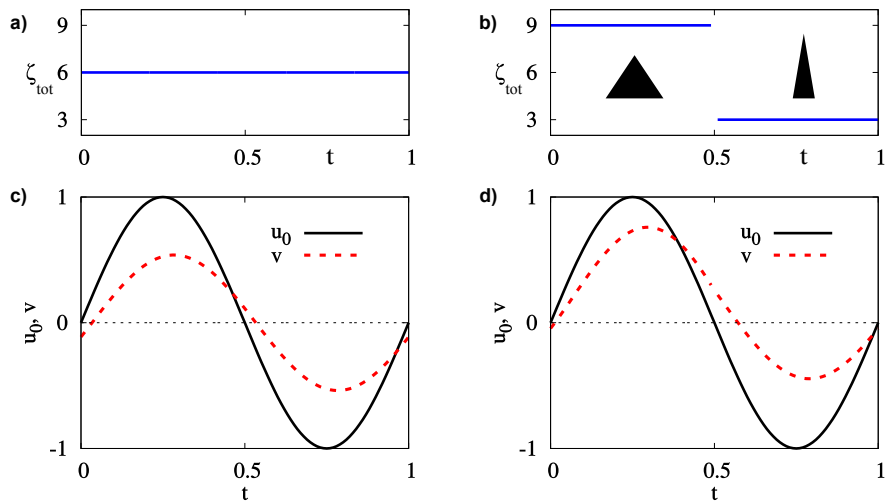}
\end{center}
\vspace{-0.7cm}
\caption{ We consider a particle of  mass  $M$  different from the  surrounding  liquid of mass  $M_f$ for the identical volume. 
We assume a sinusoidal fluid velocity $u_0(t)$ given by equation (\ref{eq_flow}) for $\varepsilon=0$, see also solid lines in c) and d).
If the shape and the Stokes drag of the particle is constant in time, the particle 
velocity  is also a sinusoidal, but has a smaller amplitude and follows with a small phase shift (dashed line in figure c)).
For a different shape and a Stokes drag in both half periods, i.e. $\zeta_1\not =\zeta_2$, 
the particle velocity is also differently, as indicated by the 
dashed line in figure d). This leads to
a different mean velocity of the particle in each half cycle. Therefore, the actuation step is different in each half cycle which results in
a net particle actuation. Parameters: $U=1$, $T=1$, $M_f=1$, $M=2$.}
\label{fig_u_v}
\end{figure*}

In the range $\frac{M}{\zeta_{tot}}\omega \gtrapprox 1$ the particle's inertia becomes important and it can not follow the 
liquid velocity, which results in $A<U$, $\phi<0$. This lag behind of the particle can be used to achieve a 
non-vanishing mean velocity: If the shape of the deformable particle and therefore  the  drag is different 
in each half cycle of the shaking, as indicated in 
figure \ref{fig_u_v}b), the delay of the particle with respect to the fluid is different in each half cycle. 
This difference may finally lead to
a net motion of the particle with respect to the fluid. 
Since the liquid does not move in the mean, this relative net motion
results in an absolute particle actuation.

In order to gain further analytical insight,  we consider an  asymmetric, i.e. anisotropic, deformable particle as illustrated 
in figure \ref{fig_u_v}b). We assume a fixed  shape and therefore a fixed 
Stokes drag during each half period as  described by
\begin{equation}
\zeta_{tot}(t) =
\left\{
\begin{array}{l}
\zeta_1 \mbox{ at } 0<t<\frac{T}{2}\,,\\
\zeta_2 \mbox{ at } \frac{T}{2}<t<T \,,
\end{array}\right.
\label{eq_zeta_t}
\end{equation}
and continued analogously in the following periods. These two different constant values of the Stokes drag just mimic the essence of the different time-dependent shapes and  Stokes drags of the particles 
sketched in figure \ref{Sketchsink}. 
Numerical results of the full equations, i.e. that include the deformations of the particles, are given in the next section.

For a sinusoidal liquid velocity the particle velocities in both half periods are 
\begin{eqnarray}
  v_{1,2}(t)  =  C_{1,2} e^{-\frac{\zeta_{1,2}}{M} t}+A_{1,2}\sin\left(\omega t+\phi_{1,2}\right)\,,
 \label{eq_v_zeta_change}
 \end{eqnarray}
 whereby $A_i$ and $\phi_i$ are calculated as given in equations (\ref{eqn_def_A}) and (\ref{eqn_def_phi}) but with the according value of $\zeta_{tot}(t)$.
Due to the periodic liquid motion, the boundary conditions for the particle velocities are
\begin{equation}
 v_1(0)=v_2(T),\qquad v_1\left(\frac{T}{2}\right)=v_2\left(\frac{T}{2}\right)\,.
\end{equation}
This allow the  determination of the constants $C_{1,2}$ as
\begin{eqnarray}
 C_1 & = & -U\omega\Gamma  \frac{(\omega^2M^2-\zeta_1\zeta_2)(e^{-\frac{2\zeta_2\pi}{\omega M}}+e^{-\frac{\zeta_2\pi}{\omega M}})}
 { e^{-\frac{\pi(\zeta_1+2\zeta_2)}{\omega M}} - e^{-\frac{\zeta_2\pi}{\omega M}}}\,,\\
 C_2 & = &  -U\omega\Gamma  \frac{(\omega^2M^2-\zeta_1\zeta_2)( 1 + e^{-\frac{\zeta_1\pi}{\omega M}})}
 { e^{-\frac{\pi(\zeta_1+2\zeta_2)}{\omega M}} - e^{-\frac{\zeta_2\pi}{\omega M}}}
\end{eqnarray}
with the abbreviation 
\begin{equation}
\Gamma = \frac{(\zeta_1-\zeta_2)(M-M_f) }{(\omega^2M^2+\zeta_2^2) (\omega^2M^2+\zeta_1^2)}\,.
\label{eq_gamma}
\end{equation}
The mean velocity of the particle is then given by
\begin{eqnarray}
  v_n &=& \frac{\int_0^Tv(t)dt}{T}\nonumber \\
  &=& \Gamma\ \frac{U \omega^2M}{2\zeta_1\zeta_2\pi} 
 \left[(\zeta_1+\zeta_2)(\zeta_1\zeta_2+\omega^2M^2) \right.\nonumber\\
 &&\qquad + \left.(\zeta_1-\zeta_2)(\omega^2M^2-\zeta_1\zeta_2)\frac{\exp\frac{\zeta_2\pi}{\omega M}-\exp\frac{\zeta_1\pi}{\omega M}}{\exp\frac{\pi(\zeta_1+\zeta_2)}{\omega M}-1}\right]\,. 
 \label{eq_vm_ana}
\end{eqnarray}
The equations (\ref{eq_gamma}) and (\ref{eq_vm_ana}) allows to discuss the requirements of a particle actuation, i.e. a non-vanishing  mean velocity $v_n$ and its direction. All factors of the product in equation (\ref{eq_vm_ana}) are positive (with assuming $U>0$), except $\Gamma$, so that the direction of the motion is determined by $\Gamma$ (this is shown in \ref{sign_vn_SI}). Furthermore, to achieve a mean velocity the factor $\Gamma$ must not be zero. This means firstly that the mass density of the particle and the surrounding fluid must differ, i.e. $M\not =M_f$. In addition 
the drag coefficients in both half cycles have to be different, i.e., $\zeta_1\not = \zeta_2$.
This is further illustrated in figure \ref{fig_u_v}.

 For an equal mass density of the particle and the liquid, $M=M_f$, the particle follows the fluid motion instantaneously
 and the mean velocity vanishes.
For $M\not =M_f$ but identical drag coefficients  in both half periods, as in figure \ref{fig_u_v}a), the 
fluid velocity ${\bf u}_0$ and the particle velocity ${\bf v}(t)$ are both sinusoidal as indicated in figure   \ref{fig_u_v}c). Both velocities 
have a different 
amplitude and there is a relative phase shift, but there is again no net progress of the particle.

If  the shape and  drag coefficients of the anisotropic particle in figure \ref{fig_u_v}b)
are different in  both half cycles of the shaking,  i.e. $\zeta_1\not = \zeta_2$,
then one has a non-symmetric velocity ${\bf v}(t)$ of the particle as shown in figure \ref{fig_u_v}d). This asymmetry of ${\bf v}(t)$
causes a  net progress of the particle per cycle. 

The net progress of a deformable particle  depends strongly on the relaxation time $\tau_v$. For a small frequency,
i.e., $\tau_v \omega\ll1$, one obtains only a small actuation 
because the particle follows the liquid's motion nearly instantaneously, i.e. ${\bf v}(t)\approx{\bf u}_0 $. This means $v_n \rightarrow0$ for $\omega\rightarrow 0$ which follows
also with equation (\ref{eq_vm_ana}).

The direction is determined by $\Gamma$ (eq. (\ref{eq_gamma})) if $U>0$ is assumed. This means the direction is given by the half period with the higher drag coefficient, i.e. by $\zeta_1$ or $\zeta_2$. It is also important if the particle has a higher or lower density than the surrounding fluid which determines if $M_f<M$ or $M_f>M$. With $M_f<M$ the particle lags behind the flow and with a lighter particle $M_f<M$ the opposite is the case.

Hence the requirements of the particle actuation can be discussed with equation (\ref{eq_vm_ana}). Note that it
takes the particle inertia into account, i.e. the terms proportional to the mass of the particle in eqs. (\ref{eq_m_bead_1}), (\ref{eq_newton}) or (\ref{max_riley_approx}) are important. The fluid Reynolds number is not important for the actuation, because the Stokes friction in equation (\ref{max_riley_approx}) occurs also at a low Reynolds number.

A time-dependence of the drag coefficient can be achieved with a soft particle in a shaken fluid. The difference in the drag coefficient in both half periods, i.e. $\zeta(t)\neq\zeta(t+T/2)$, (as sketched in figure \ref{fig_u_v}b)) can be achieved with an asymmetric particle in a sinusoidal shaken fluid. In case of a symmetric soft particle a different shape and therefore a different drag coefficient of the particle in each half cycle can be achieved 
by a non-symmetric periodic fluid velocity ${\bf u}_0(t)$ with $\varepsilon \not =0$ in equation (\ref{eq_flow}).
This is further exemplified in the next section.

To compare the approximation in this section for ${\bf v}$ and the results from simulations in the next section, 
gravity must be taken into account. Gravity leads approximately to the additional contribution 
\begin{equation}
 v_s=\frac{g}{2}\left(\frac{M_f-M}{\zeta_1}+\frac{M_f-M}{\zeta_2}\right)\,
\end{equation}
to the mean actuation velocity, cf. equation (\ref{eq_vm_ana}).\\
 
\section{Numerical Results}\label{results}
In this section, we explore  numerically   the inertia driven dynamics and locomotion of
 four soft particles in a shaken liquid, which are sketched in figure  \ref{Sketchsink}.
The selected numerical simulations are guided by the results of the previous section i.e.,
particle locomotion is expected in parameter ranges with different
mass densities of  the particles and the liquid  and the Stokes drag of a particle is 
is unequal during each half of a shaking period $T$. 
Such a time-dependent Stokes drag can be realized by asymmetric particles but also with symmetric ones.

Firstly, the dynamics of the asymmetric particles is investigated for the sinusoidal shaking velocity ${\bf u}_0$ in equation (\ref{eq_flow}) with $\varepsilon=0$. We show simulations of the bead-spring tetrahedron in section \ref{tetradyn} and compare 
the results with a more realistic Janus capsule in section \ref{janusdyn}.
The intrinsic anisotropic of both particles causes  
different deformations and Stokes drags
during each half period of the sinusoidal velocity ${\bf u}_0$. 

Secondly, symmetric particles are investigated. To achieve different deformations and Stokes drags during the two halves of the shaking period for these particles as well, we utilize a non-symmetric shaking velocity in equation  (\ref{eq_flow}) with $\varepsilon \not =0$.
We discuss the dynamics of a symmetric bead-spring ring and show that a symmetric capsule behaves similar to the ring in section \ref{capsuledyn}.

All particles are soft particles with a different mass density than the liquid. They are deformed in shaken liquids, which is taken into account in the numerical simulations.
 Hence, besides the velocity relaxation time $\tau_v$ (cf. equations (\ref{def_tau_v}) and (\ref{def_tau_v_caps})) considered in sec. \ref{sec_explanation}, also the shape relaxation time $\tau_k$ and especially the ratio $T/\tau_k$ are important. The shape relaxation time is given by the time the particles needs to relax to their equilibrium shape after a deformation. 
To determine the order of the relaxation-time scale, we use as an estimate for
the bead spring models 
\begin{equation}
\tau_k\approx\sqrt{\frac{m}{k}}
\label{def_tau_k}
\end{equation}
with the spring constant $k$, cf. equation (\ref{eq_m}), and the bead mass $m$
and for the capsules 
\begin{equation}
\tau_k\approx\sqrt{\frac{\rho_{capsule} V}{G}}
\label{def_tau_k_caps}
\end{equation}
with the capsule volume $V$ and the surface shear elastic modulus $G$.

\subsection{Actuation of a tetrahedron in a sinusoidally shaken liquid \label{tetradyn}}
\begin{figure}[h]
\begin{center}
\includegraphics[width=0.34\textwidth]{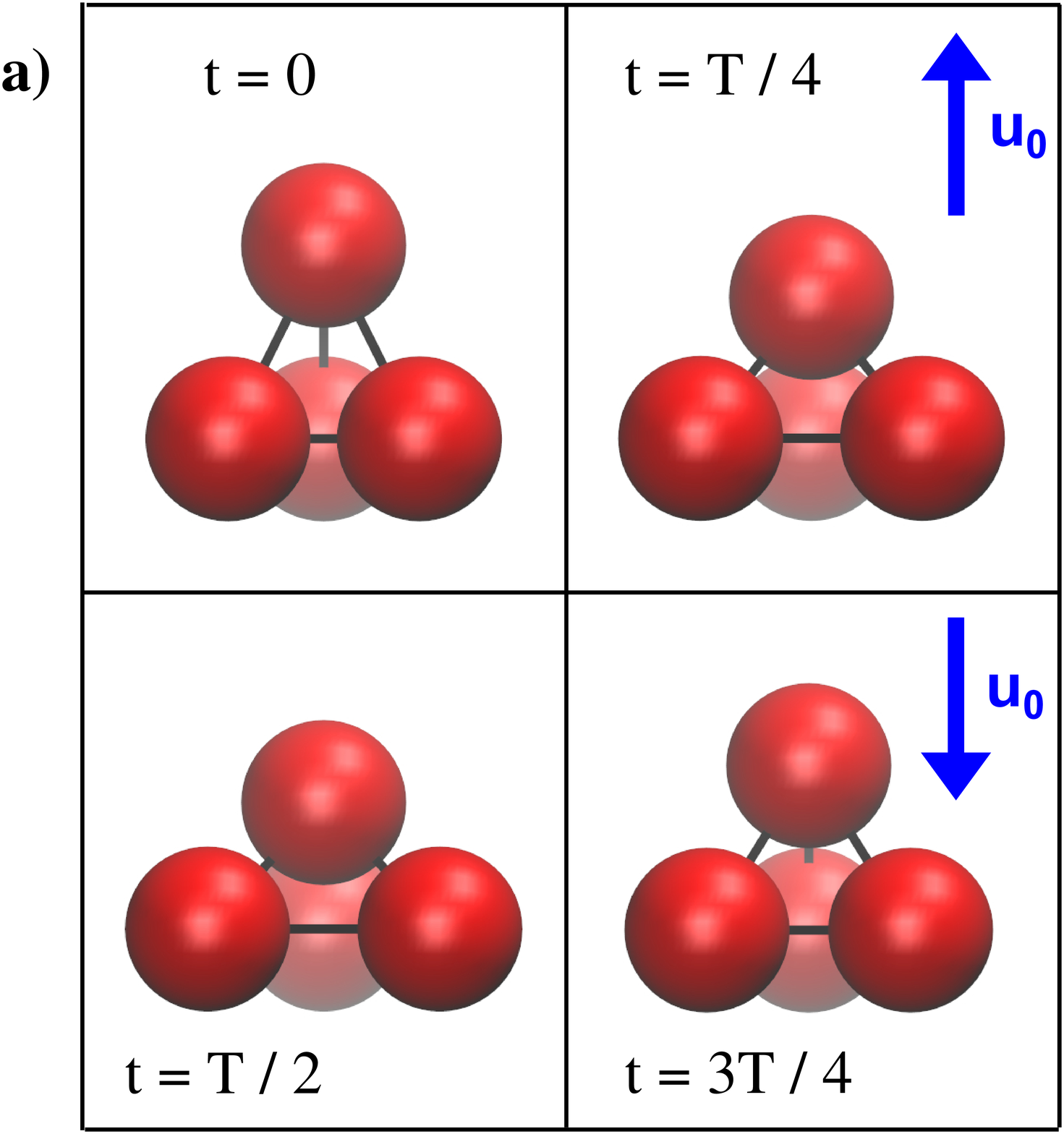}
\quad 
\includegraphics[width=0.48\columnwidth]{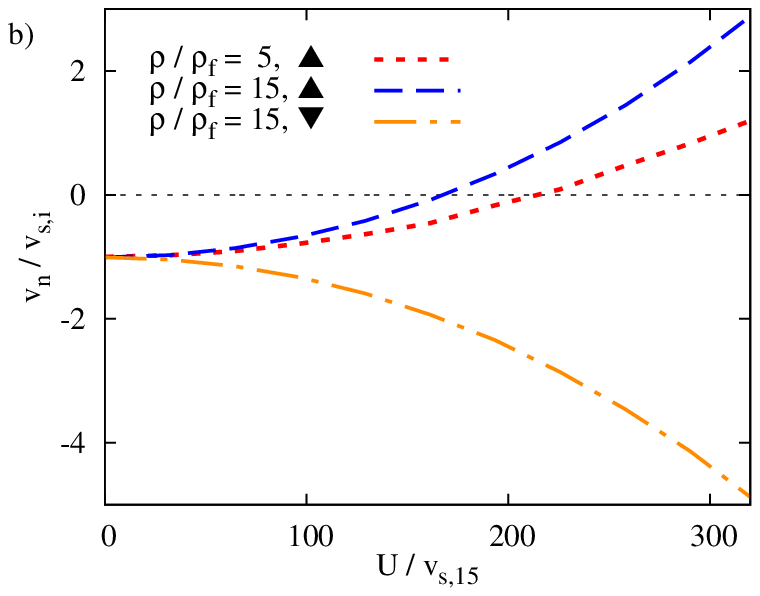}
\end{center}
\vspace{-0.7cm}
\caption{a) shows four snapshots of a deformable, upward oriented tetrahedron($\blacktriangle$)
 during one period $T$ in a sinusoidally shaken fluid.
In  b) the mean propulsion velocity $v_n$ of the tetrahedron is given for two ratios
 between the mass density of the beads and  of the fluid, i.e. 
for $\rho/\rho_f=5,15$. $v_n$ is given in units of the related sedimentation 
velocities $v_{s,i}$, respectively.
The $\blacktriangle$-tetrahedron
outweighs gravity for $\rho/ \rho_f=15$ in the range $U/v_{s,15}  \gtrsim 160$ and rises in the shaken liquid.
For $\rho/\rho_f=5$ the $\blacktriangle$-tetrahedron rises in the range $U/v_{s,15}  \gtrsim 210$. The sedimentation velocity of
the $\blacktriangledown$-tetrahedron is enhanced by liquid shaking as indicated by the dash-dotted line. Parameters: see section \ref{models_oseen}.
}
\label{fig_tetraact_1}
\end{figure}

We investigate at first the motion of asymmetric particles in a sinusoidally shaken fluid. 
We begin with the simple bead-spring tetrahedron.
Two orientations of the bead-spring tetrahedron in a vertically shaken  fluid are investigated, one 
with a corner upward ($\blacktriangle$), cf. figure \ref{fig_tetraact_1}a), and one with a  corner downward  
($\blacktriangledown$). These positions are stable against a rotational perturbation.
Figure  \ref{fig_tetraact_1}a)  shows the $\blacktriangle$-tetrahedron  
at four deformations during one period $T$ of a sinusoidally shaken liquid.

The center of mass of the tetrahedron, $y_c(t)$, follows via the viscous drag
the oscillatory motion of the shaken liquid.  Moreover, $y_c(t)$ exhibits besides an
oscillatory motion also a  mean  net propulsion  as indicated in figure \ref{Sketchsink}a).
The  resulting mean velocity $v_n$ of the center of mass, which is studied in the following, 
is determined by fitting a straight line to $y_c(t)$  over a sufficient number of periods after a transient phase. 
The  parameters for the numerical studies are given in section \ref{models_oseen}.
We give the mean velocity $v_n$ and the amplitude $U$  of the  shaking velocity in  units of
the sedimentation velocity (absolute value) denoted by $v_{s,r}$, whereby the index $r$ indicates 
the ratio of the density of the tetrahedron and the fluid,  $\rho/\rho_f=r$. 
The sedimentation velocities $v_{s,5}=8.9\cdot10^{-3}$ and $v_{s,15}=3.1\cdot10^{-2}$ (absolute values) are determined without a shaking of the liquid (pure sedimentation).

In figure \ref{fig_tetraact_1}b)
we show the mean velocity $v_n$  
of the tetrahedra in the gravitational field as function of the amplitude $U$. For the
$\blacktriangle$-tetrahedron for two ratios $\rho/\rho_f=5,15$ and for the 
$\blacktriangledown$-tetrahedron for $\rho/\rho_f=15$.
The sedimentation velocity $v_{s,r}$
and $v_n$ increase with the density ratio $\rho/\rho_f$. 
For $\blacktriangle$-tetrahedra  the mean velocity $v_n$ becomes positive for
 $\rho/\rho_f=15$  beyond 
$ U/v_{s,15} \gtrsim 160$ and for 
$\rho/\rho_f=5$ beyond 
$ U/v_{s,15} \gtrsim 210$. In both  ranges the locomotion of a tetrahedron 
 outweighs the downward oriented gravitation and heavy particles can be elevated.
 Therefore,  for smaller mass differences between 
soft particles and the liquid this locomotion mechanism becomes less effective 
and a higher velocity amplitude $U$ is required to outweigh
gravitation. A downward orientated shaken heavy
tetrahedron ($\blacktriangledown$) will sediment faster than without shaking. 
Furthermore, a buoyant particle with $\rho/\rho_f=1$ follows the oscillatory liquid motion
and its mean velocity $v_n$ vanishes in agreement with the reasoning given in the previous section.
The inertial actuation is also found for tetrahedra lighter than the liquid, i.e. $\rho/\rho_f<1$. 
Note that the dependence on the initial condition can be avoided by an asymmetric mass distribution 
of the beads, because this leads to a reorientation of the tetrahedron. 
For example with one bead lighter than the other three beads, 
the lighter bead will point upwards after a certain time. The inertia driven actuation of such a tetrahedron is discussed in \ref{tetra_SI}.

\begin{figure*}[htb]
\begin{center}
\includegraphics[width=0.45\textwidth]{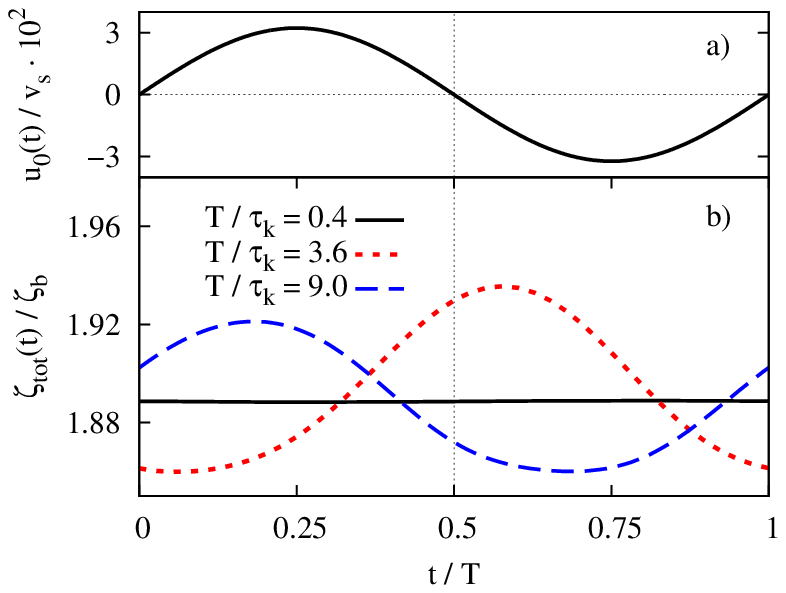}
\qquad \includegraphics[width=0.44\textwidth]{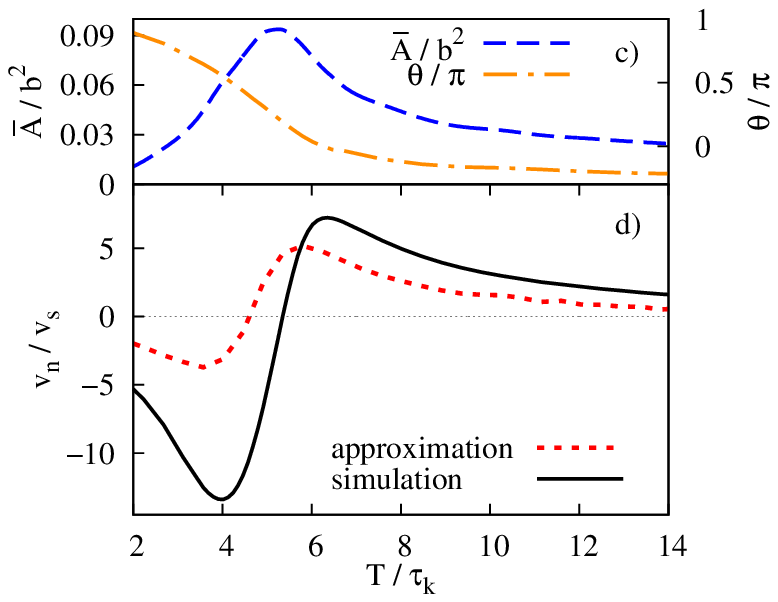}
\end{center}
\vspace{-0.7cm}
\caption{a) shows the shaking velocity $u_0(t)$  with $\varepsilon=0$ in equation (\ref{eq_flow})
in units of the sedimentation velocity $v_s$ 
of an upward oriented tetrahedron   ($\blacktriangle$)  with $\rho/\rho_f=15$.
The deformation of a tetrahedron, cf. figure \ref{fig_tetraact_1}a), is accompanied
by a time-dependent Stokes drag $\zeta_{tot}$ as shown in b) (in units of $\zeta_b=6\pi\eta a$)
for three different ratios  $T/\tau_k$.
c) shows the time-dependent deviation  $\bar A$
from the mean area $A_0$ of the lower triangle 
of a  $\blacktriangle$-tetrahedron,  as defined in equation (\ref{tetraground}),
as well as the phase shift 
$\Theta/\pi$ between the velocity $u_0(t)$ and the deformation. 
In  d) the dependence of the
mean velocity $v_n/v_s$ is given as a function of $T/\tau_k$ for the velocity amplitude $U/v_s \simeq 32$.
The dashed line is obtained by equation  (\ref{eq_vm_ana}) with $\zeta_{1,2}$.
Parameters: $k=30000$ and those given in section \ref{models_oseen}.}
\label{fig_tetraact_2}
\end{figure*}

The  mean velocity $v_n$   depends also on ratio between the shaking period $T$ and the bead-spring relaxation time $\tau_k$ (cf. equation (\ref{def_tau_k})). This dependence  is shown in figure \ref{fig_tetraact_2}d) for a $\blacktriangle$-tetrahedron. This figure shows in part b) also the
time-dependence of the drag coefficient of the tetrahedron, $\zeta_{tot}(t)$ (cf. \ref{models_oseen_SI} ), 
which is caused by the time-dependent deformation.
Thus, in addition the deformation amplitude  $\bar A$ of the bottom triangle of the tetrahedron with area
\begin{equation}
\label{tetraground}
A(t)=\bar{A}\sin\left(\frac{2\pi}{T}t-\Theta\right)+A_0\,,
\label{eq_def_bar_A}
\end{equation}
is given in figure \ref{fig_tetraact_2}c). 
Also the phase shift $\Theta$ of the deformation and the flow is shown. The area A(t) is determined by a fit to the data.

If $T$ is considerably smaller than the
relaxation time $\tau_k$, the  deformation of the tetrahedron cannot follow the liquid oscillation
and remains small, as indicated for the deformation amplitude ${\bar A}$ in figure \ref{fig_tetraact_2}c). Consistently, 
the drag coefficient $\zeta_{tot}$  is nearly constant as indicated for $T/\tau_k=0.4$
in figure \ref{fig_tetraact_2}b).
In this case  particles just sediment in a shaken liquid.
For larger $T$ the tetrahedron becomes deformed during liquid shaking 
and the drag coefficient  $\zeta_{tot}(t)$ shows similar as $u_0(t)$ a sinusoidal time-dependence
as indicated for $T/\tau_k=3.6$ in figure \ref{fig_tetraact_2}b).
However, for such short shaking periods 
 the tetrahedron deformation  can still not follow the liquid oscillation and 
$\zeta_{tot}$ is nearly in antiphase to $u_0(t)$ as indicated by in figure \ref{fig_tetraact_2}b) and in figure \ref{fig_tetraact_2}c).
Due to this phase shift for $T/\tau_k=3.6$ the Stokes drag in figure
\ref{fig_tetraact_2}b) is larger during the downward liquid motion with $u_0(t)<0$ than during its upward motion.
Therefore, the inertia induced locomotion is downward oriented for $T/\tau_k=3.6$ as also indicated 
in figure \ref{fig_tetraact_2}d) for the whole range $T/\tau_k \lesssim 5.7$.
For $T/\tau_k=9$ beyond the maximum of $v_n/v_s$ in \ref{fig_tetraact_2}d)
the deformation of the tetrahedron follows $u_0(t)$ more closely with a smaller phase difference $\Theta$
and the drag  is slightly larger during the upward motion, cf. figure \ref{fig_tetraact_2}b).
In this case and in the range $T/\tau_k \gtrsim 5.7$  the locomotion 
mechanism points into the opposite direction to the gravitation
and can even outweigh gravitation for $U/v_s \simeq 32$, i.e., $v_n/v_s$ becomes positive.
 $v_n/v_s$  remains positive up to about $T/\tau_k \sim 27$ and beyond this ratio the tetrahedron sinks again due to the gravitation. This shows that indeed the time-dependent Stokes drag coefficient is a cause of the non-zero mean velocity. Note hereby that the model neglects the advective terms of the Navier-Stokes equation, i.e. they are not important for the actuation.
 
Besides the shape relaxation time also the velocity relaxation time  $\tau_v$ (cf. equation (\ref{def_tau_v})) plays a role as stated in section \ref{sec_explanation}. We have chosen similar values of 
$\tau_v\approx 0.07$ and $\tau_k\approx 0.03$. 
The period $T$ is in the range $1\lesssim T/\tau_v\lesssim 6$, so that the particle's inertia is significant.

\begin{figure}[htb]
\begin{center}
\includegraphics[width=0.45\columnwidth]{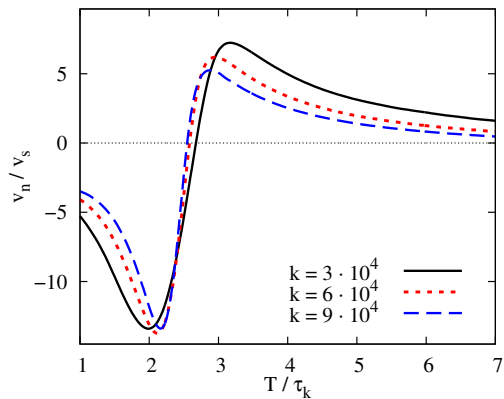}
\end{center}
\vspace{-0.7cm}
\caption{The mean locomotion velocity $v_n$ of the $\blacktriangle$-tetrahedron, cf. figure \ref{fig_tetraact_1}a), 
is given as a function of $T/\tau_k$ for three different values of the spring stiffness $k$
in equation (\ref{eq_m}) with $\rho/\rho_f=15$. The minima of the curve occur at similar values of $T/\tau_k\approx 2$ and the maxima at $T/\tau_k\approx3$ despite different values of $k$. }
\label{fig_vm_k}
\end{figure}

One can also use the approximate mean velocity  in equation (\ref{eq_vm_ana}) by selecting the drag coefficient
 $\zeta_{tot}$ from simulations of the tetrahedron. We use the maximal drag during each half period for $\zeta_{1,2}$. 
The resulting dependence of $v_n/v_s$ is indicated in figure \ref{fig_tetraact_2}d).
This result confirms that the 
approximate approach presented in section \ref{sec_explanation} 
covers the essential inertia driven locomotion mechanism considered in this work.

In figure \ref{fig_vm_k} the dependence of $v_n/v_s$ on the ratio $T/\tau_k$ is shown for different values
of the spring constant $k$ of the tetrahedron. The extrema and the zero of $v_n/v_s$ 
are located at similar values of $T/\tau_k$. Moreover, the
magnitudes of the minima and maxima of $v_n/v_s$ differ only slightly for different values of $k$.
This emphasis again the importance of the ratio between shaking period and the particle's relaxation time $\tau_k$
\subsection{Actuation of a Janus capsule in a sinusoidally shaken liquid\label{janusdyn}}

\begin{figure}[htb]
\begin{center}
\includegraphics[width=0.3\columnwidth]{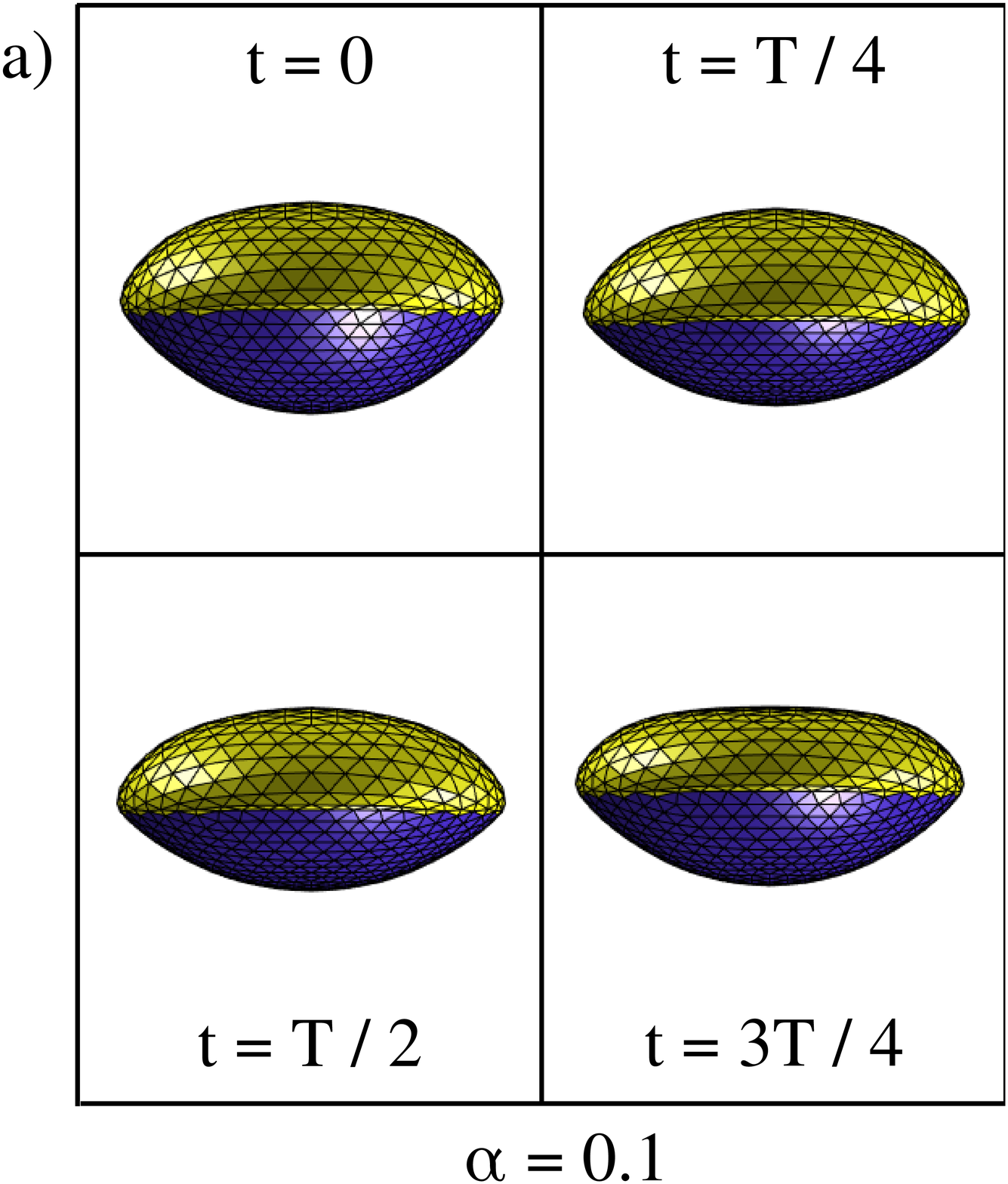}
\qquad \quad
\includegraphics[width=0.475\columnwidth]{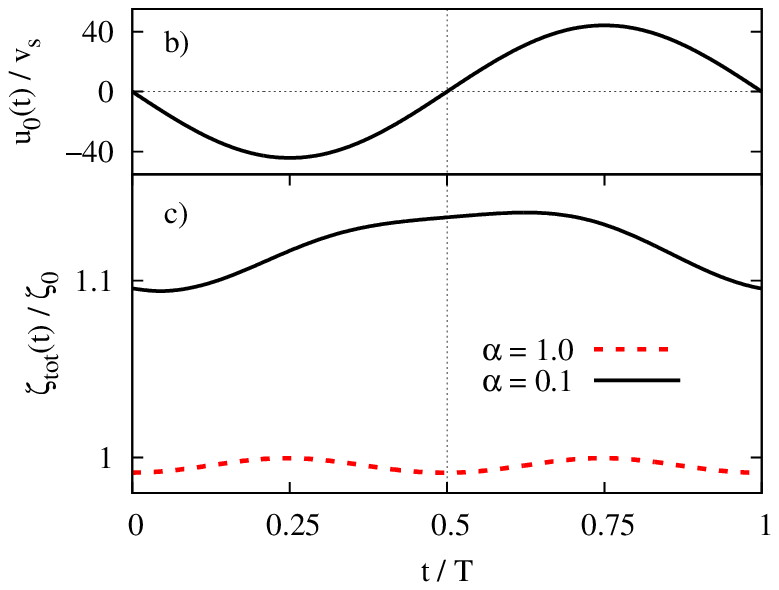}
\end{center}
\caption{a) shows four snapshots of a deformable, 
with the soft side upward oriented Janus capsule
($\blacktriangle$) during a  period $T$
in a shaken fluid with the elasticity ratio $\alpha=0.1$. Figure
b) shows the sinusoidal shaking velocity $u_0(t)/v_s$ in units of the sedimentation velocity $v_s$
of the Janus capsule.
The lower part in b) shows the Stokes drag $\zeta_{tot}(t)$ in units 
of the Stokes drag $\zeta_0=6\pi\eta R$ of the undeformed capsule for the elasticity ratio $\alpha=0.1$ (solid line) 
and the symmetric capsule with $\alpha=1$ (dashed line). 
}
\label{fig_asym_caps_1}
\end{figure}

With a Janus capsule that is composed of two parts of different elasticity we consider in this section 
a realistic soft anisotropic particle. The
four snapshots shown in  \ref{fig_asym_caps_1}a) 
 highlight the different deformations 
 during a sinusoidal shaking cycle $T$.
We investigate two orientations of the asymmetric Janus particle in the shaken liquid: 
One with the soft half on top as in figure \ref{fig_asym_caps_1}a) (upward oriented Janus capsule $\blacktriangle$),
 or with the soft part at the bottom  ($\blacktriangledown$). These orientations are stable against a rotational perturbation.

The capsule simulations are performed with the LBM and, besides the parameters given in section \ref{models_LBM}, the following values are used: radius of the capsule $R=10\mu\mbox{m}$,
$\rho_{Janus}=2 \rho_{fluid}$, $G^{(0)}=3.95\cdot 10^{-3}\mbox{kg}/\mbox{s}^2$ 
and $\kappa_c^{(0)}=3,77\cdot 10^{-13}\mbox{kg m}^2/\mbox{s}^2$.
For the elastic properties of the second half of the capsule we set 
$\kappa_c^{(var)}=\alpha \kappa^{(0)}$ and $G^{(var)}=\alpha G^{(0)}$
with an elasticity ratio  $\alpha=0.1$. 
This results in the two ratios $T/\tau_k\approx 2$ and $T/\tau_v\approx 2$ (cf. equations (\ref{def_tau_v_caps}) and (\ref{def_tau_k_caps}), 
determined with $ G^{(0)}$), 
which ensure that the Janus capsule is deformed during the shaking of the liquid and that the inertia of the capsule is significant.

For Janus capsules neither the deformation nor 
the Stokes drag $\zeta_{tot}$ has a symmetry, i.e. $\zeta_{tot}(t)\neq\zeta_{tot}(t+T/2)$,
as indicted for $\alpha=0.1$ by the snapshots in figure \ref{fig_asym_caps_1}a) 
and in figure \ref{fig_asym_caps_1}c), respectively. 
Therefore the time-dependence of $y_c(t)$ 
for a $\blacktriangle$ Janus capsule in  figure \ref{Sketchsink}b) 
displays the  inertia induced locomotion (cf. figure \ref{Sketchsink}b) ). This is not the case for the symmetric 
capsule with $\alpha=1.0$: The Stokes drag $\zeta_{tot}(t)$ given in figure \ref{fig_asym_caps_1}c)
has the symmetry $\zeta_{tot}(t)=\zeta_{tot}(t+T/2)$. Consequently, the  symmetric capsule just sinks 
in a sinusoidally shaken fluid in the presence of gravitation.

\begin{figure}[t]
\begin{center}
\includegraphics[width=0.45\columnwidth]{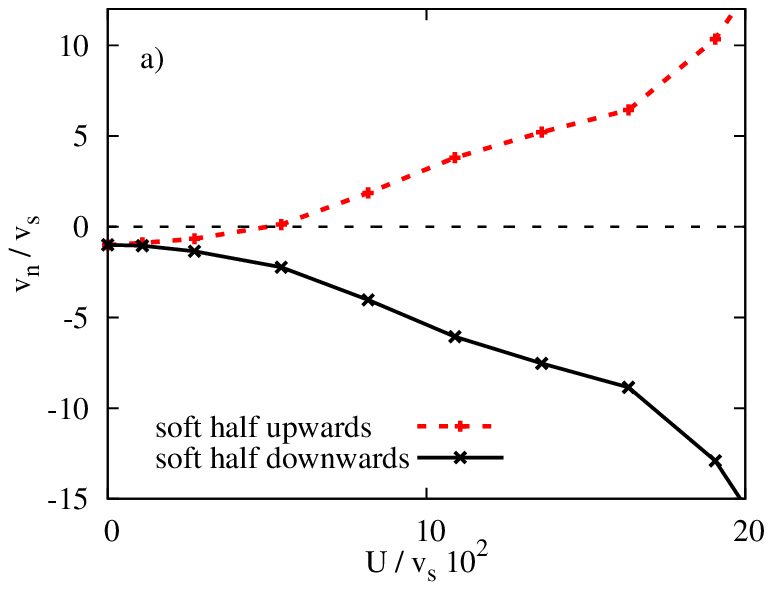}
\quad
\includegraphics[width=0.45\columnwidth]{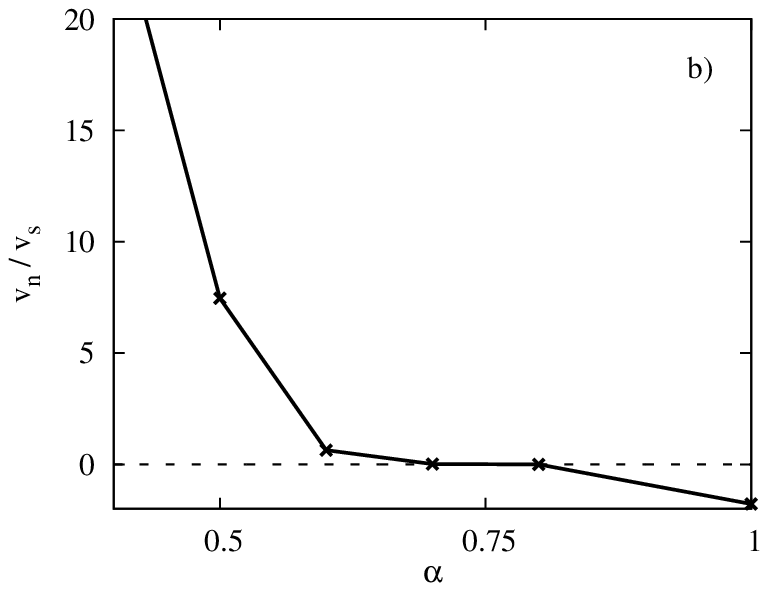}
\end{center}
\vspace{-0.7cm}
\caption{The mean actuation velocity $v_n$ of a Janus capsule
is given as a function of the shaking-velocity amplitude $U$ in equation (\ref{eq_flow})
 with the  soft side either upwards  ($\blacktriangle$) or downwards ($\blacktriangledown$). 
In the  $\blacktriangle$-case the capsule locomotion outweighs gravitation if  $U\gtrsim 500\,v_s$ and $v_n$ becomes positive.
In the  $\blacktriangledown$-case the shaking of the liquid enhances the sinking velocity.
b) shows  $v_n/v_s$ for the $\blacktriangle$-capsule as function of the elasticity ratio $\alpha$. 
At $\alpha=1$ the particle sinks.
By enhancing the asymmetry (decreasing $\alpha$) the capsule locomotion outweighs
gravitation for $\alpha \lesssim 0.8$.
%
}
\label{fig_asym_caps_vm_alpha}
\end{figure}

The mean velocity of the Janus capsule $v_n/v_s$ is shown
in figure \ref{fig_asym_caps_vm_alpha} as function of the velocity amplitude $U$.
The sedimentation velocity 
of the  $\blacktriangledown$ Janus capsule 
is enhanced by the oscillatory fluid motion 
as shown by the lower curve in \ref{fig_asym_caps_vm_alpha}a).
For a velocity amplitude $U\gtrsim 500\,v_s$ (with $v_s\approx 0.15 \frac{\mbox{mm}}{\mbox{s}}$)
the locomotion of the $\blacktriangle$
capsule outweighs gravitation and moves upward, i.e., $v_n>0$.
This is indicated by the dashed curve in \ref{fig_asym_caps_vm_alpha}a). As for the 
tetrahedron the locomotion increases with the difference  between the mass density of the  capsule and
the liquid. This also means, the critical amplitude $U$ to outweigh gravitation  is reduced by
increasing the ratio $\rho_{capsule}/\rho_{liquid}$.
The mean locomotion velocity $v_n/v_s$ for an upward oriented Janus capsule in a gravitational field 
is also shown as function of elasticity ratio $\alpha$ in figure \ref{fig_asym_caps_vm_alpha}b).
This graph shows  that the inertia induced locomotion increases with increasing elastic asymmetry 
(i.e., decreasing $\alpha$) and outweighs  in the range $\alpha \lesssim0.8$ gravitation for the given parameters.
The symmetric capsule with $\alpha=1.0$ just sinks in the mean.

In figure \ref{fig_asym_caps_vm_alpha}a) the Reynolds number Re in LBM simulations is
finite with \hbox{$0\leq$ Re $\lesssim3$}  and an
 inertia induced capsule locomotion is found at small and intermediate values of the Reynolds number (and also beyond this values). 
 The qualitative behavior of this capsule locomotion 
 is similar as for the tetrahedron in the limit of vanishing Reynolds number. The reason is that the locomotion of the particles is driven by the inertia of the particles and the time-dependent stokes drag, which is both included in the model of the bead-spring tetrahedron.

\subsection{Bead-spring ring in a non-symmetrically shaken liquid \label{ringdyn}}

In this and the following section, we explore the conditions for which also common symmetric soft micro-particles
behave in  shaken liquids as passive microswimmers.
We begin with a symmetric bead-spring ring as sketched in figure \ref{Sketchsink}.
The parameters used in simulations are given in section \ref{models_oseen} and 
the velocities are given  in units of the sedimentation velocity $v_s=0.031$ (determined without shaking of the liquid).

\begin{figure}[htb]
\begin{center}
\includegraphics[width=0.30\columnwidth]{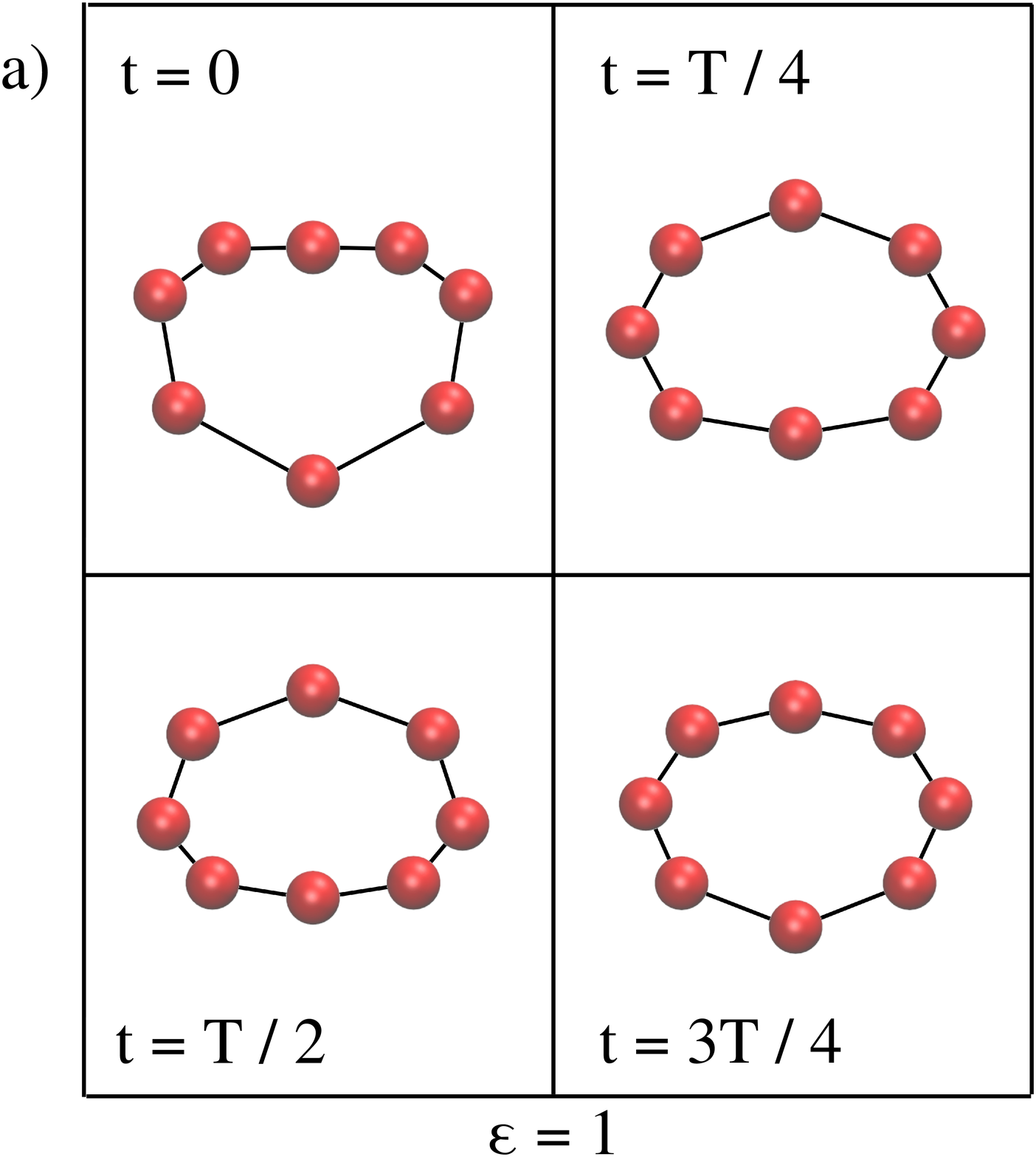}
\quad \includegraphics[width=0.45\columnwidth]{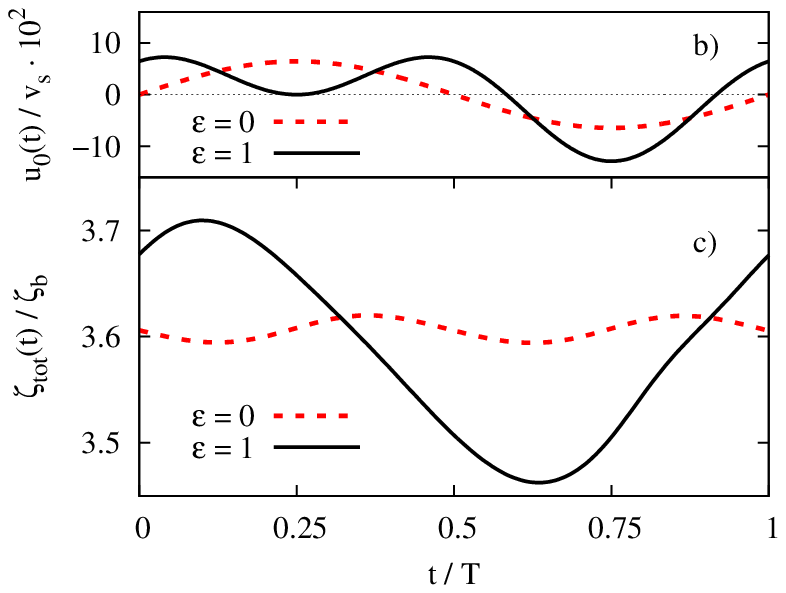}
\end{center}
\caption{a) shows four snapshots of the bead-spring ring  during one period $T$ 
of the liquid velocity as shown by the solid line
in  b) and as given by equation (\ref{eq_flow}) for $\varepsilon=1$. 
In this case the drag coefficient $\zeta_{tot}(t)$ (in units of $\zeta_b=6\pi\eta a$) is different in both half periods as shown
by the solid line in c), i.e., $\zeta_{tot}(t+T/2)\not =\zeta_{tot}(t)$. This causes
a finite mean actuation velocity $v_n$.
For $\varepsilon=0$ the shaking is sinusoidal, cf. dashed line in b), the
drag coefficient is the same in both half periods of the shaking, i.e.,  $\zeta_{tot}(t)=\zeta_{tot}(t+T/2)$, and $v_n=0$.}
\label{fig_ring_zeta_t}
\end{figure}
Figure \ref{fig_ring_zeta_t}a) shows four snapshots of a  bead-spring ring during one  
period $T$ of a non-symmetric  shaking velocity $u_0(t)$ given by equation (\ref{eq_flow}) 
and  as shown in figure  \ref{fig_ring_zeta_t}b) for $\varepsilon=1$.
For a sinusoidally 
shaken liquid with $\varepsilon=0$ and $u_0(t)=-u_0(t+T/2)$ the drag coefficient $\zeta_{tot}(t)$ (cf. \ref{models_oseen_SI} ) 
is the same in both half periods  with $\zeta_{tot}(t)=\zeta_{tot}(t+T/2)$, as indicated in figure \ref{fig_ring_zeta_t}c). 
In this case the ring exhibits no net actuation 
and sinks in the gravitational field.
For a  non-symmetric periodic shaking velocity 
with $\varepsilon \neq 0$  and $u_0(t)\neq-u_0(t+T/2)$ the drag coefficient of the ring
is different in both half periods as shown for $\varepsilon=1$ in \ref{fig_ring_zeta_t}c). This leads to the passive swimming as shown in figure \ref{Sketchsink}c).

\begin{figure}[htb]
\begin{center}
\includegraphics[width=0.5\columnwidth]{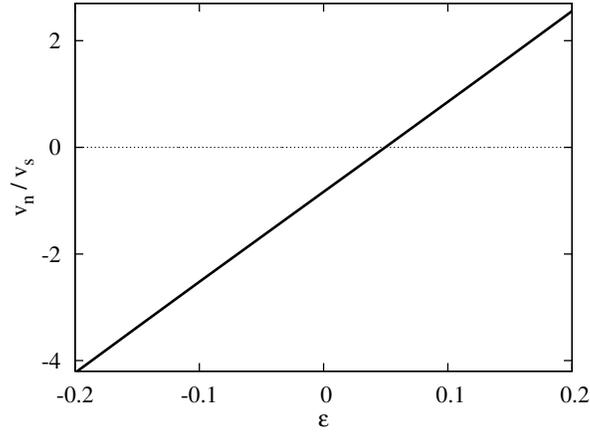}
\end{center}
\vspace{-0.7cm}
\caption{
The mean actuation velocity  $v_n$ varies linearly with the modulation parameter $\varepsilon$ of the shaking velocity
in equation (\ref{eq_flow}). 
For sufficiently positive values  $\varepsilon \gtrsim 0.05$ the liquid shaking outweighs gravitation
and $v_n$ becomes positive.
For sinusoidal shaking with $\varepsilon=0$ the particle sinks due to gravity 
and at negative values of $\varepsilon$ the inertial actuation leads to an enhanced sedimentation velocity.}
\label{fig_ring_vm_eps}
\end{figure}

Figure \ref{fig_ring_vm_eps} shows the mean velocity  $v_n$ of the bead-spring ring 
as a function of $\varepsilon$.  
For  $\varepsilon \gtrsim 0.05$ the upward directed, inertia induced actuation is sufficiently strong 
 to outweigh gravitation and $v_n$ becomes positive.  For $\varepsilon<0$ liquid shaking enhances the
sedimentation velocity. 
Thus the  sign of $\varepsilon$ determines the direction of the inertia induced actuation of the semiflexible ring.

\begin{figure}[htb]
\begin{center}
\includegraphics[width=0.45\columnwidth]{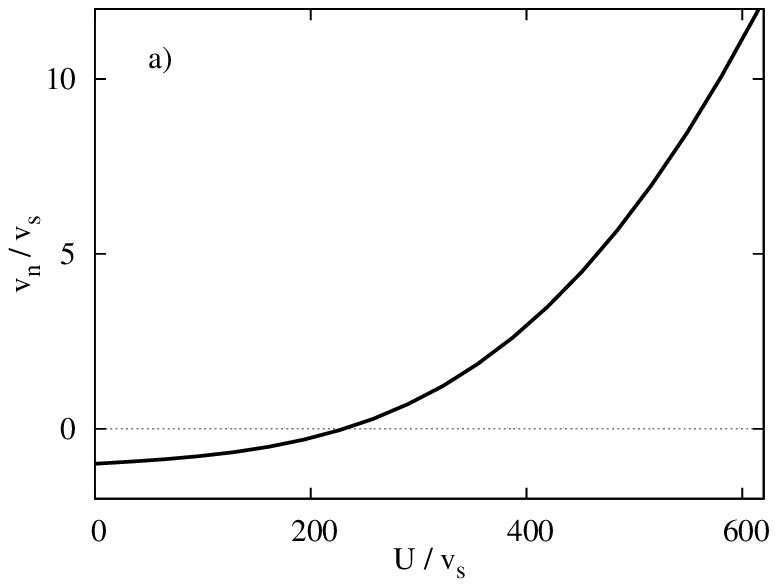}
\qquad 
\includegraphics[width=0.45\columnwidth]{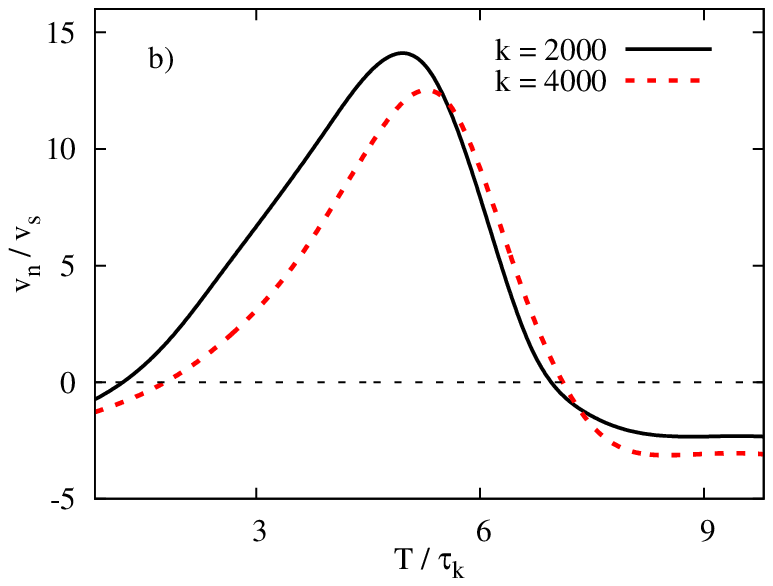}
\end{center}
\vspace{-0.7cm}
\caption{a) shows the mean propulsion velocity $v_n$ as a function of the shaking amplitude $U$. 
For  positive $\varepsilon$ the  actuation is upward directed and for $\varepsilon=1$ it outweighs gravitation
in the range  $U\gtrsim225 v_s$, i.e.,   $v_n$  becomes positive. 
b) shows the mean velocity of the ring as a function of the ratio ${T}/{\tau_k}$ for two different 
values of the spring stiffness $k$. At small values of ${T}/{\tau_k}$ 
the ring can not follow the liquid motion, it is not deformed and  just sinking. 
At intermediate values ${T}/{\tau_k}$ the mean velocity becomes positive for both values of $k$ with a
maximum in the range ${T}/{\tau_k}\approx 5$. For longer shaking periods the ring sinks again. 
}
\label{fig_ring_vm_U}
\end{figure}

The mean velocity $v_n$ is given as function of the amplitude $U$ of the shaking velocity in figure \ref{fig_ring_vm_U}. 
Without shaking at $U=0$ the ring sinks. With increasing values of $U$ and $\varepsilon=1$ the sinking velocity slows down until it 
turns over to an upward motion at larger values $U\gtrsim225 v_s$.

The mean velocity $v_n$ depends also  on the ratio between the shaking period and the relaxation time 
 $T/\tau_k$ as shown in \fref{fig_ring_vm_U}b) 
for two values of the spring stiffness $k$. 
At  small values  of  ${T}/{\tau_k}\lesssim0.4$ the ring sinks because 
the shaking period is too small to cause  sufficient deformations and differences 
between the Stokes drags during the two half periods.
For longer periods $T$ and intermediate values of $T/\tau_k$ the acceleration induced shape
and Stokes drag  changes of the ring become sufficiently strong to outweigh gravitation. 
For both spring constants $k$ 
the mean velocity becomes positive in a wide range and reaches its maximum at 
a value of ${T}/{\tau_k}\approx 5$ due to the large deformation, as indicated 
in figure \ref{fig_ring_vm_U}b). 
At higher values of ${T}/{\tau_k}$ the deformation 
becomes smaller and therefore  the ring sinks again due to  gravitation. 
The values of the shape relaxation time $\tau_k\approx 0.08$ (cf. equation (\ref{def_tau_k})) 
and the velocity relaxation time $\tau_v\approx0.07$ (cf. equation (\ref{def_tau_v})) are comparable, 
so that $T/\tau_v$ is in a range where the particle's inertia is important.

\subsection{Actuation of a symmetric capsule in a non-symmetrically shaken liquid \label{capsuledyn}}
In the previous section we demonstrated that a symmetric, semiflexible bead-spring ring 
is actuated in a liquid that is non-symmetrically shaken with $u_0(t)\neq-u_0(t+T/2)$.  
This is also the case  for a realistic symmetric capsule as we show 
by LBM simulations in this section. Besides the parameters given in section \ref{models_LBM}, the following ones are used: 
$R=10\mu\mbox{m}$, $\rho_{capsule} = 2 \rho_{fluid} = 2000 \frac{\mbox{kg}}{\mbox{m}^3}$, 
$k_v = 2.78\cdot10^5\frac{\mbox{kg}} {\mbox{s}^2\mbox{m}}$, $G=7.89\cdot 10^{-4}\frac{\mbox{kg}}{\mbox{s}^2}$, $\varepsilon=-1$
and $\kappa_c=1.51\cdot 10^{-14}\frac{\mbox{kg m}^2}{\mbox{s}^2}$. The shaking period $T$ is chosen so, 
that the capsule's inertia is significant and the capsule is sufficiently deformed:
$T/\tau_v\approx 2$ and $T/\tau_k\approx 0.9$ (cf. equations (\ref{def_tau_v_caps}) and (\ref{def_tau_k_caps})).

\begin{figure}[h]
\begin{center}
\includegraphics[width=0.295\columnwidth]{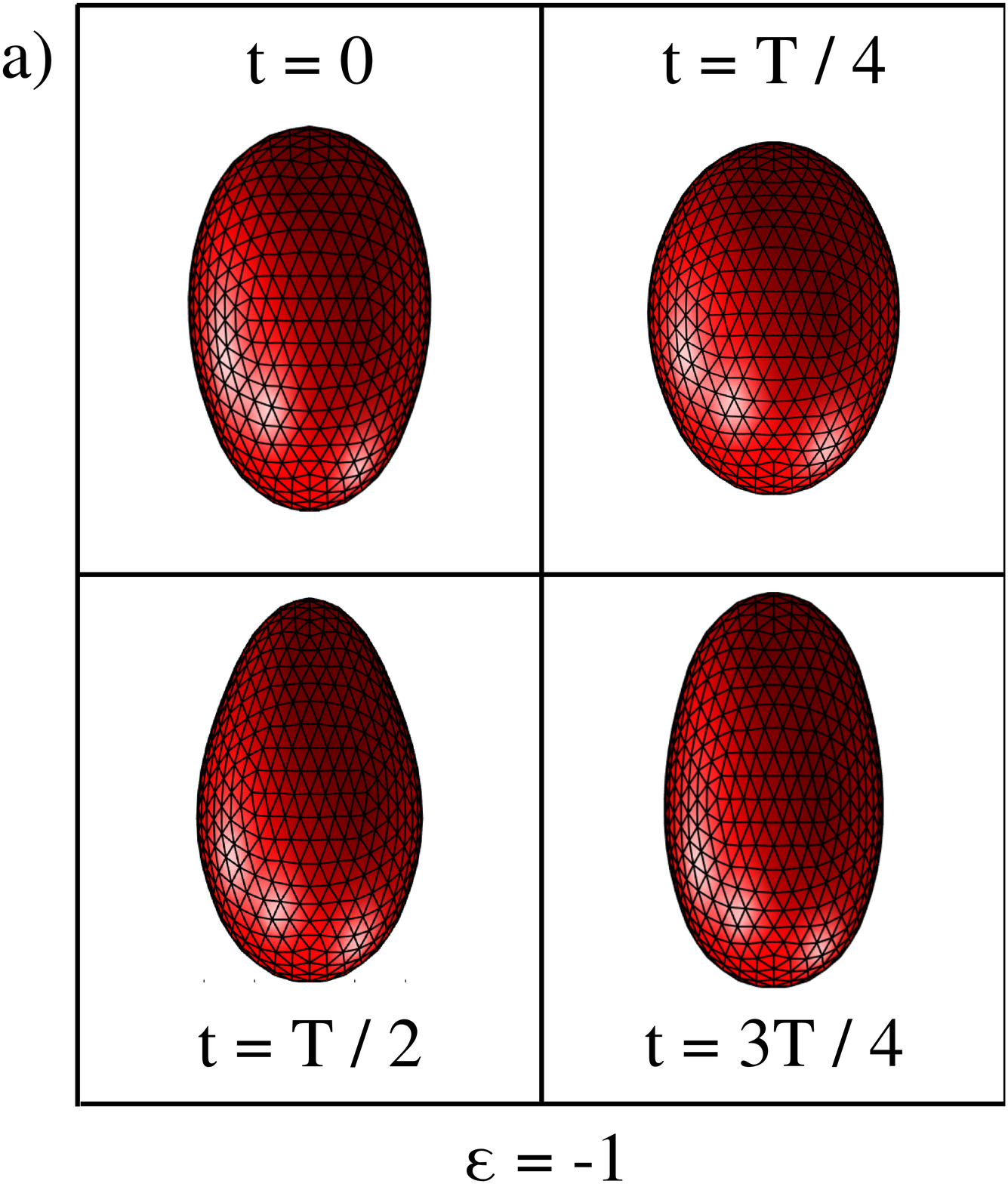}
\quad
\includegraphics[width=0.455\columnwidth]{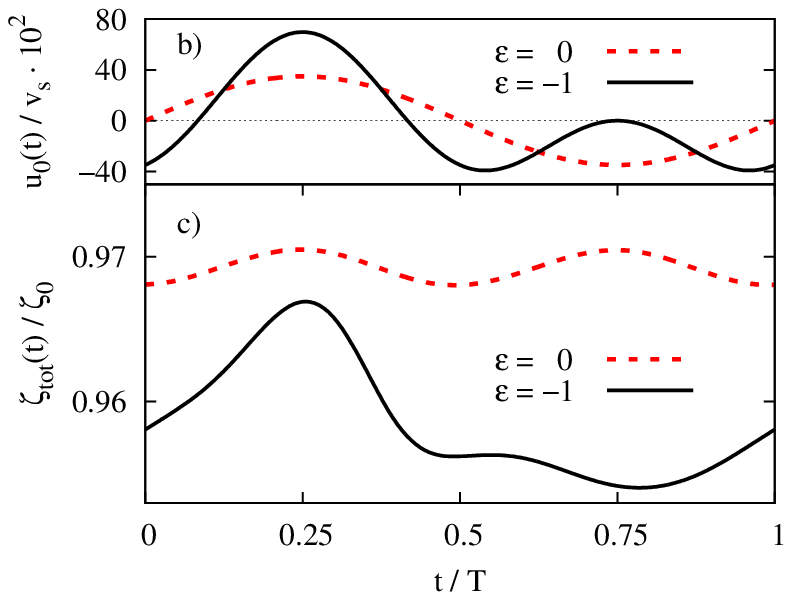}
\end{center}
\caption{a) shows the capsule's shape at different times 
in a shaken liquid with a  non-symmetric
liquid velocity, i.e., $u_0(t+T/2)\not =u_0(t)$,
as shown by the solid line in b) and as given by equation (\ref{eq_flow}) for $\varepsilon=-1.0$. In
c) the drag coefficient $\zeta_{tot}(t)$ of the capsule is shown in units of $\zeta_0=6\pi\eta R$.
In a sinusoidally shaken liquid, cf. dashed line in b), $\zeta_{tot}(t)$ is identical
in both half periods. For a non-symmetrically  shaken liquid also $\zeta_{tot}(t)$ is non-symmetric,
cf. solid line and $\varepsilon=-1$, as well as the capsule shapes in a).}
\label{fig_caps_zeta_t}
\end{figure}
\begin{figure}[htb]
\begin{center}
\includegraphics[width=0.45\columnwidth]{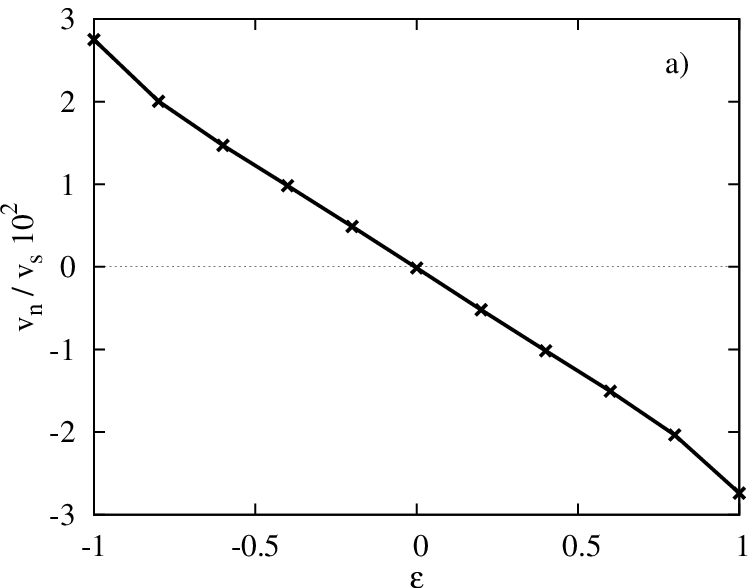}
\includegraphics[width=0.45\columnwidth]{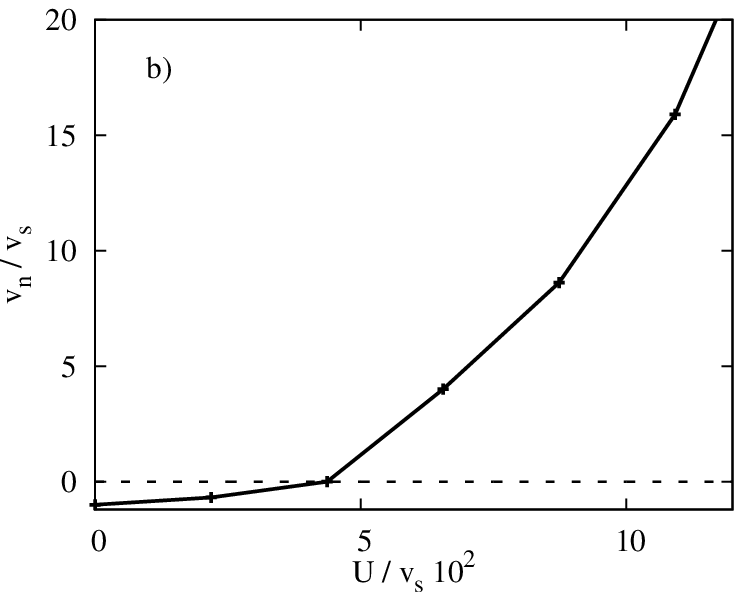}
\end{center}
\vspace{-0.7cm}
\caption{a) shows the mean-propulsion  velocity $v_n$  of a homogeneous capsule 
as a function of the asymmetry parameter $\varepsilon$  of the shaking velocity in equation (\ref{eq_flow}).
For a sinusoidal shaking with $\varepsilon=0$ the capsule sinks. 
At sufficiently negative values  $\varepsilon<0$ the capsule moves upwards and for $\varepsilon>0$ downwards. 
This allows to control the direction of $v_n$ via the time-dependence of liquid shaking.
In b) $v_n$ is given as function of the amplitude $U$ of the shaking velocity. 
At low values of $U$ the shaking effect is weak and the capsule sinks. 
In the range $U\geq 450v_s$ the actuation induced by the shaking is stronger than the gravity and the capsule 
moves upwards.}
\label{fig_caps_vm_eps_U}
\end{figure}

In figure \ref{fig_caps_zeta_t}a) the shape of the capsule is shown during one period $T$  of the non-symmetric shaking velocity with
$u_0(t+T/2)\not = u_0(t)$. 
For a sinusoidal shaking as displayed in figure \ref{fig_caps_zeta_t}b), 
i.e.,  $\varepsilon=0$ and $u_0(t)=-u_0(t+T/2)$, 
the capsule's drag coefficient $\zeta_{tot}(t)$ (cf. \ref{models_oseen_SI} )
shown in figure \ref{fig_caps_zeta_t}c) is 
the same during both half periods of the shaking with $\zeta_{tot}(t)=\zeta_{tot}(t+T/2)$.
In this case there is no mean actuation and the capsule just sediments due to gravity.
If the liquid is  shaken non-symmetrically  with $\varepsilon\not =0$ the drag coefficient 
differs in both half periods, i.e.,  $\zeta_{tot}(t)\neq\zeta_{tot}(t+T/2)$. 
In this case the capsule is actuated by liquid shaking.

Figure \ref{fig_caps_vm_eps_U}a) shows how the mean velocity $v_n$ of the capsule depends on the 
asymmetry parameter $\varepsilon$ of the shaking velocity. 
At sufficient negative values of $\varepsilon\lesssim-0.01$  the upward oriented actuation 
 overcomes gravity and we find a positive mean velocity $v_n$. 
Positive values of $\varepsilon$ enhance the sedimentation. 
Thus the direction of the mean capsule actuation can be controlled via the asymmetry parameter 
$\varepsilon$ of the shaking velocity. 
Note that the mean velocity induced by the shaking also depends on the period $T$.

Besides the asymmetry $\varepsilon$ also a sufficiently high amplitude $U$ of the shaking velocity is required to overcome gravity. 
Figure \ref{fig_caps_vm_eps_U}b) displays the mean velocity $v_n$ as a function of the amplitude $U$: 
At low values of $U$  the capsule sinks due to the gravity. For the chosen parameters one finds with $U=0$ 
the sedimentation velocity  $v_s=0.19\frac{\mbox{mm}}{\mbox{s}}$.
For $U\gtrsim450v_s$ the mean velocity $v_n$ induced by liquid shaking is stronger 
than sedimentation and the capsule moves upwards for $\varepsilon=-1.0$. 
The Reynolds number used in figure \ref{fig_caps_vm_eps_U}b) is $0<$ Re $\lesssim2$. 
Hence,  the inertia induced actuation  effect is found at small as well as at intermediate values of Re. 
The qualitative results are comparable to those found for the  ring in the previous section 
in the limit of a vanishing Reynolds number, compare e.g., figure \ref{fig_ring_vm_U}a) and figure \ref{fig_caps_vm_eps_U}b). The reason is that the requirements of the particle locomotion are the inertia of the particle and the time-dependent stokes drag, which is both included in the model of the bead-spring ring.

%
\section{Summary and conclusions}\label{conclusion}
We investigated a new kind of microswimmers, so-called passive swimmers. 
These microswimmers are soft particles with a mass density different from the liquid, 
which are driven by an oscillating background flow or a shaking of the liquid.

Previous studies focused on the propulsion of intrinsically asymmetric soft particles in sinusoidal 
liquid motion  \cite{Kanso:2016.1,Ishikawa:2018.1}. With our extension to 
  soft bead-spring tetrahedrons and to asymmetric, soft Janus capsules,
we show that the inertia driven propulsion mechanism can even outweigh gravity. 
Moreover, we  show that this novel inertia driven passive swimming mechanism 
works  for the  wider class of symmetric soft particles, such as capsules.

 By a semi-analytical model calculation we cover the essential properties 
 of the inertia driven propulsion mechanism
 in liquids shaken periodically with the velocity ${\bf u}_0(t+T)={\bf u}_0(t)$.
 It shows the following requirements: First, the mass densities of the particles and the liquid must be different. Secondly, the Stokes drag during both periods of 
 the shaking with different directions must differ (e.g., due to a deformation). Thirdly, the shaking period $T$ has to be chosen in the order of magnitude of the relaxation time that the particles needs to adjust to the liquid velocity.
The essential difference in the drag coefficient during both half periods is achieved by the asymmetry of the particle.

  We suggest that this asymmetry can also be  achieved by a  non-symmetric shaking velocity with ${\bf u}_0(t+T/2) \not ={\bf u}_0(t)$,
 as given for instance by equation (\ref{eq_flow}),  instead of 
 the intrinsic particle asymmetry. 
 Such a non-symmetric liquid shaking
 leads to a non-reciprocal particle deformation and  Stokes drag.

This qualitative reasoning and the  analytical  considerations are verified and supported by  simulations. We use symmetric and asymmetric
bead spring models and complementary Lattice Boltzmann Simulations of realistic soft symmetric capsules  and asymmetric Janus capsules.
Asymmetric particles in a sinusoidally shaken fluid have two stable orientations and they exhibit
therefore two  directions of passive swimming, depending on the initial orientation. In contrast, for the wider class
of symmetric particles in non-symmetrically shaken liquids 
the propulsion direction is determined by the shaking. Therefore the swimming direction can be selected by the engineered time-dependence of liquid shaking.

 To provide examples of achievable  propulsion velocities for  symmetric and  asymmetric Janus capsules 
 we chose a realistic capsule size of about $10\,\mu$m and a stiffness of $8\cdot10^{-4} ~\frac{\mbox{N}}{\mbox{m}}$, which fits the 
values of common capsules \cite{Chen:2015.1,Sun:2015.1,Amstad:2017.1}. 
A higher mass density for capsules than for the  liquid can be achieved if  salt is dissolved in 
the liquid inside the capsule \cite{KOROZNIKOVA:2008.1}, whereby water with dissolved 
salt can reach densities up to three times higher than pure water (without salt).
Here we chose the mass density ratio $\rho_{caps}/\rho_{liquid}=2.0$ and the shaking frequency 10 kHz (see e.g.,
 \cite{Wurmus:1995,Nabavi:2009,Roberts:2002.1})
 of the order of the inverse of velocity relaxation time of about $44 \mu\mbox{s}$.
 For this choice of parameters and a maximal amplitude 
 0.5 $\frac{\mbox{m}}{\mbox{s}}$ of the liquid velocity
 one obtains for  a symmetric capsule  with  in Lattice Boltzmann simulations 
 an upward swim velocity of about  57 $\frac{\mbox{mm}}{\mbox{s}}$.
 For a Janus capsule one obtains for shaking-velocity amplitude  0.3 $\frac{\mbox{m}}{s}$
 an upward swim velocity of about 15 $\frac{\mbox{mm}}{\mbox{s}}$.

Besides the possibility to engineer passive swimmers, the described effects have further applications: 
 The inertia induced actuation may be exploited for separating particles with respect to their
different mass and different elasticity (deformability).
The separation of two  kinds of soft particles with a different stiffness
is achieved by choosing a  shaking period that fits the shape relaxation time of one 
type of particles but not of the others. 
In this case one particle type is stronger actuated and can be accumulated for instance near one container wall.
An example are biological cells. They have often a different density than water \cite{Milo:2016.1}
or other carrier liquids. In addition the
 stiffness of cells is often  an indicator 
of their health status \cite{Karimi:2013.1,Cross:2007.1,Guck:2005.1}. 
In this case healthy cells may be  separated for instance from malignant cells by non-symmetric
liquid shaking.  Our insights about inertia driven particle  propulsion might also have impact 
on further systems studied at finite values of the Reynolds number \cite{Sorokin:2012.1,Loewen:2018.1}.

\section{Acknowledgments}
 We acknowledge the support by the French-German University (Grant CFDA-Q1-14, program ``Living fluids'').

\appendix
\section{Dynamic Oseen tensor and drag coefficient} \label{models_oseen_SI} 
\paragraph{Dynamic Oseen tensor.}
The liquid velocity ${\bf u}({\bf r}_i)$ at the particle positions ${\bf r}_i$
includes the imposed homogeneous flow ${\bf u}_0$ as described by equation (\ref{eq_flow})
and the flow disturbances caused by differences between the particle velocities ${\bf v}_j$ and the liquid velocity  ${\bf u}_0({\bf r}_j)$.
For this 
we use the general solution of the linear part of the Navier-Stokes-equation 
$\rho_f\frac{\partial \mathbf{u}}{\partial t}=\eta \Delta \mathbf{u}-\nabla p+\mathbf{f}(\mathbf{r},t)$ for an arbitrary point-like force acting 
on the fluid. The solution of this problem with a point force $\mathbf{f}(\mathbf{r},t)=\mathbf{F}(t)\delta(r-r')$ is given by \cite{Ignacio:1995}
\begin{eqnarray}
 \mathbf{u}_{\delta}(\mathbf{r})&=&\frac{1}{\rho_f}\int_0^t dt'\ \mathbf{H}({\mathbf{r}-\mathbf{r}'},t')\mathbf{F}(t')\,,\\
 \mathbf{H}(\mathbf{r})&=&p(r,t)\mathbf{1}-q(r,t)\frac{\mathbf{r\otimes r}}{r^2}\,,\\
 p(r,t)&=&\left(1+\frac{2\nu t}{r^2}\right)f(r,t)-\frac{g(r,t)}{r^2}\,, \label{eq_dyn_ossen_p}\\
 q(r,t)&=&\left(1+\frac{6\nu t}{r^2}\right)f(r,t)-\frac{3g(r,t)}{r^2}\,, \label{eq_dyn_ossen_q}\\
 f(r,t)&=&\frac{1}{(4\pi\nu t)^{3/2}}\exp\left[\frac{-r^2}{4\nu t}\right]\,,\\
 g(r,t)&=&\frac{1}{4\pi r}\Phi\left[\frac{r}{(4\nu t)^{1/2}}\right]\,,
\end{eqnarray}
with $\nu=\frac{\eta}{\rho}$, the error function $\Phi$, the unit matrix $\mathbf{1}$ and the dyadic product $\otimes$.
This allows to calculate  the liquid velocity  at a bead  position $\mathbf{u}_i=\mathbf{u}(\mathbf{r}_i)$,
\begin{equation}
\mathbf{u}_i=\mathbf{u}_0(t)-\frac{1}{\rho_f}\sum\limits_{j\neq i}\int\limits_0^t dt'\ \mathbf{H}_{i,j}(t')\cdot\mathbf{F}_j^{(1)}(t')\\
 \end{equation}
with $\mathbf{H}_{i,j}(t)=\mathbf{H}({\mathbf{r}_i-\mathbf{r}_j},t)$.
This velocity is composed of the homogeneous background flow and the liquid velocity changes caused
by the differences between the particle velocities and the flow ${\bf u}_0$, which are induced
by the forces ${\bf F}_j^{(1)}$ given in section \ref{models_oseen}.

\paragraph{Determination of the drag coefficient.}
To calculate the drag $\zeta_{tot}(t)$, we follow the procedure given in \cite{Dhont:96,DoiEd,Leal:2007}. 
For this, we use the positions of the beads/nodes on the particle surface obtained by simulations. The drag at time $t$ is determined by assuming a fixed shape which implies a constant velocity $\mathbf{v}_i=\mathbf{v} = v \hat{\mathbf{e}}_y$ on each bead/node. We calculate the forces via
\begin{equation}
 \mathbf{v}_i=\sum_j\mathbf{H}_{i,j}\cdot\mathbf{F}^{(P)}_j,
\end{equation}
where 
\begin{equation}
\mathbf{H}_{i,j} = \left\{ \begin{array}{l} \mathbf{O}(\mathbf{r}_i-\mathbf{r}_j),\\\frac 1 \zeta_b \mathbf{1} \end{array} \right.
\end{equation}
is the mobility matrix including the hydrodynamic interaction between particle $\mathbf{r}_i$ and $\mathbf{r}_j$ described by the Oseen tensor $\mathbf{O}(\mathbf{r})=\frac{1}{8\pi\eta r}\left(\mathbf{1}+\frac{\mathbf{r}\otimes\mathbf{r}}{r^2}\right)$. The drag finally follows with
\begin{equation}
	\mathbf{F}_{tot} = \sum_j \mathbf{F}_j = \zeta_{tot} \mathbf{v}.
\end{equation}
The values $\zeta_{1}$ and $\zeta_2$ (used in equation (\ref{eq_zeta_t})) are chosen as the maximal value during the first or the second halve period, respectively.\\


\section{Tetrahedron consisting of beads with different mass}\label{tetra_SI}
Here we investigate the effects of the mass inhomogeneity on the propulsion velocity of  a tetrahedron.
If all beads of a tetrahedron have the same mass density, the upward oriented tetrahedron ($\blacktriangle$)
and the downward oriented one ($\blacktriangledown$) are both stable. 
By changing the mass density of  one of the four beads then one of  both orientations 
with respect to  the gravitational field is preferred, similar 
as in reference \cite{Ishikawa:2018.1}.
For example, if the tetrahedron sinks (without liquid shaking) 
the lighter bead points upwards after a certain time. For this orientation we investigate 
the effect of a inhomogeneous mass density  on
the propulsion velocity.

We introduce the density ratio  $\alpha$ between one and the other three beads, i.e.,  $\rho_1=\alpha \rho_{2,3,4}$, 
and keep  the mean density $\bar \rho$ constant:
\begin{eqnarray}
 \overline{\rho}&=&\frac{1}{N}\sum\limits_{i=1}^N\rho_i \label{rho_mean}\,,\\
 \rho_1&=&\alpha\rho_2,\  \rho_2=\rho_3=\rho_4 \label{alpha}\,.
\end{eqnarray}
\begin{figure}[h]
\begin{center}
\includegraphics[width=0.45\columnwidth]{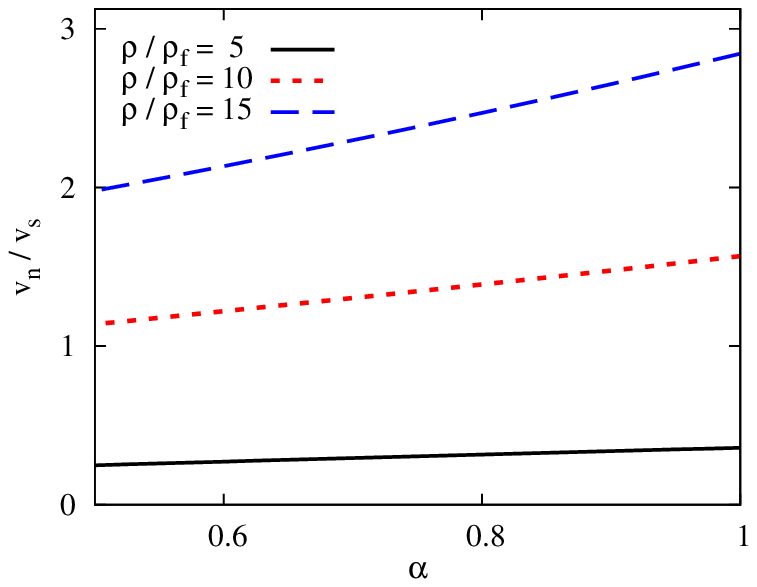}
\includegraphics[width=0.45\columnwidth]{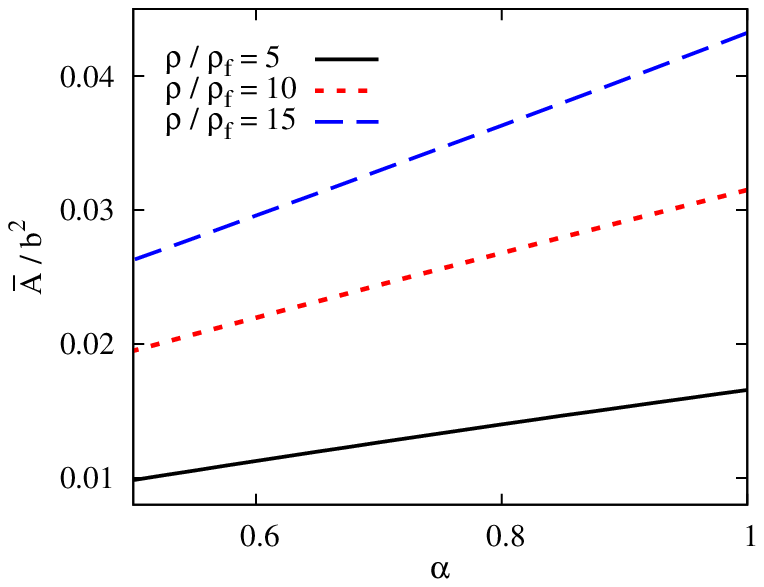}
\end{center}
\vspace{-0.7cm}
\caption{The left part shows the  mean propulsion velocity of the tetrahedron for different 
mass-density ratios $\bar\rho /\rho_f$  
as function 
of the density ratio $\alpha$
and
the right part the amplitude of the shape deformation $\bar{A}$ as defined 
in equation (\ref{eq_def_bar_A}).
}
\label{fig_vm_alpha}
\end{figure}
Figure \ref{fig_vm_alpha} shows the mean velocity $v_n$ of the tetrahedron and 
the amplitude of the shape deformation $\bar{A}$ (defined in equation (\ref{eq_def_bar_A})) 
as a function of the mass-density ratio $\alpha$ simulated with the Maxey Riley equations. The tetrahedron moves slower with an increasing difference 
of the densities of the beads, which can be explained as follows.
A lighter bead can follow the heavier ones easily and thus the lighter bead moves 
more in phase with the heavy beads than a bead of the same mass density.
This results in smaller spring deformation and  in a lower deformation amplitude $\bar{A}$, cf.
 figure \ref{fig_vm_alpha}. A smaller amplitude $\bar A$ 
leads to smaller  temporal changes 
of the drag coefficient $\zeta_{tot}$ and therefore to  slower mean velocity.

\begin{figure}[h]
\begin{center}
\includegraphics[width=0.5\columnwidth]{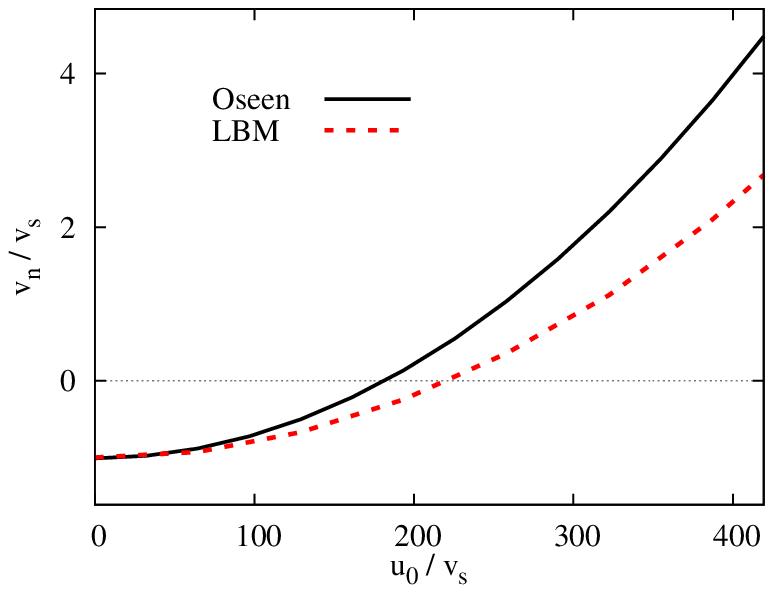}
\end{center}
\vspace{-0.7cm}
\caption{The mean propulsion velocity of a tetrahedron is shown 
for $\alpha=0.6$ as a function of the amplitude $U$ of the liquid velocity, 
obtained by  simulations of the Maxey-Riley  equations and Lattice Boltzmann simulations.
}
\label{fig_vm_vgl}
\end{figure}

So far we have used the Maxey and Riley equations and the dynamic Oseen tensor, 
i.e., we have neglected effects of a finite Reynolds number. 
Here we compare the results with  Lattice Boltzmann simulations of the full Navier-Stokes equation with the tetrahedron. 
Figure \ref{fig_vm_vgl} shows the mean velocity of a tetrahedron with $\alpha=0.6$ 
as function of the amplitude $U$ of the shaking velocity. Both methods show that the mean velocity 
increases continuously with the amplitude $U$. Furthermore both simulations show  that 
at low values of $U$ the tetrahedron sinks and above a critical value of $U$ the tetrahedron 
rises against gravity. Thus the numerical methods agree qualitatively.
This means the LBM  simulations, taking effects of a finite  Reynolds number into account,
and the  Maxey and Riley equations including the dynamic Oseen tensor in the limit  $Re=0$
describe inertia induced propulsion of the tetrahedron. 
This confirms that the mean velocity is the result of the temporal change 
of the drag coefficient $\zeta_{tot}(t)$ and a finite Reynolds number just modifies this result quantitatively.

\section{Discussion of the sign of the mean velocity}\label{sign_vn_SI}
The mean velocity $v_n$ of the particle is given by equations (\ref{eq_gamma}) and (\ref{eq_vm_ana}) in the main text as follows
\begin{eqnarray}
\Gamma &=& \frac{(\zeta_1-\zeta_2)(M-M_f) }{(\omega^2M^2+\zeta_2^2) (\omega^2M^2+\zeta_1^2)}\,,\nonumber\\
  v_n &=& \frac{\int_0^Tv(t)dt}{T}\nonumber \\
  &=& \Gamma\ \frac{U \omega^2M}{2\zeta_1\zeta_2\pi} 
 \left[(\zeta_1+\zeta_2)(\zeta_1\zeta_2+\omega^2M^2) \right.\nonumber\\
 &&\qquad + \left.(\zeta_1-\zeta_2)(\omega^2M^2-\zeta_1\zeta_2)\frac{\exp\frac{\zeta_2\pi}{\omega M}-\exp\frac{\zeta_1\pi}{\omega M}}{\exp\frac{\pi(\zeta_1+\zeta_2)}{\omega M}-1}\right]\,. \nonumber
\end{eqnarray}
We show here that the sign of the mean velocity is determined by $\Gamma$ because all other factors in the equation of $v_n$ (equation (\ref{eq_vm_ana})) are positive.

The factor $\frac{U \omega^2M}{2\zeta_1\zeta_2\pi}$ is positive because we assume $U>0$. We define furthermore 
\begin{eqnarray}
A=(\zeta_1+\zeta_2)(\zeta_1\zeta_2+\omega^2M^2)\\
B=(\zeta_1-\zeta_2)(\omega^2M^2-\zeta_1\zeta_2)\frac{\exp\frac{\zeta_2\pi}{\omega M}-\exp\frac{\zeta_1\pi}{\omega M}}{\exp\frac{\pi(\zeta_1+\zeta_2)}{\omega M}-1}
\end{eqnarray}
which leads to
\begin{equation}
  v_n  = \Gamma\ \frac{U \omega^2M}{2\zeta_1\zeta_2\pi} 
 \left[A+B\right]\,. \nonumber
\end{equation}
It is trivial that $A$ is positive. We demonstrate now $A+B>0$ by showing that $|A|>|B|$. We compare the absolute values of $A$ and by $B$ for each factor. It is clear that
\begin{eqnarray}
|\zeta_1+\zeta_2|>|\zeta_1-\zeta_2|\,, \label{C1}\\
|\zeta_1\zeta_2+\omega^2M^2|>|\omega^2M^2-\zeta_1\zeta_2|\label{C2}\,.
\end{eqnarray}
To show
\begin{equation}
1\geq\left|\frac{\exp\frac{\zeta_2\pi}{\omega M}-\exp\frac{\zeta_1\pi}{\omega M}}{\exp\frac{\pi(\zeta_1+\zeta_2)}{\omega M}-1}\right|
\end{equation}
we define 
\begin{eqnarray}
a_{1,2}&=&\frac{\zeta_{1,2}\pi}{\omega M}>0\,,\\
f(a_1,a_2)&=&\frac{\exp a_2-\exp a_1}{\exp(a_1+a_2)-1}\,.
\end{eqnarray}

The function $f$ has the following properties: it increases monotonously with $a_2$ and decreases monotonously with $a_1$. Furthermore it is
\begin{eqnarray}
f(0,a_2)=1\,,\\
\lim_{a_1\to\infty}f(a_1,a_2)=-\exp{(-a_2)}\,,\\
f(a_1,0)=-1\,,\\
\lim_{a_2\to\infty}f(a_1,a_2)=\exp{(-a_1)}\,.
\end{eqnarray}
With $a_{1,2}\geq 0$ follows
\begin{eqnarray}
-1\leq f(a_1,a_2) \leq 1\,,\\
|f(a_1,a_2)|<1\,. \label{C3}
\end{eqnarray}
The equations (\ref{C1}), (\ref{C2}), (\ref{C3}) lead to $|A|>|B|$ and with $A>0$ follows $A+B>0$. Therefore we get for eq. (equation (\ref{eq_vm_ana})

\begin{eqnarray}
  v_n  &=& \Gamma\ \underbrace{\frac{U \omega^2M}{2\zeta_1\zeta_2\pi}}_{>0}\nonumber\\
 &&
 \underbrace{\left[(\zeta_1+\zeta_2)(\zeta_1\zeta_2+\omega^2M^2)(\zeta_1-\zeta_2)(\omega^2M^2-\zeta_1\zeta_2)\frac{\exp\frac{\zeta_2\pi}{\omega M}-\exp\frac{\zeta_1\pi}{\omega M}}{\exp\frac{\pi(\zeta_1+\zeta_2)}{\omega M}-1}\right]}_{>0} \nonumber
\end{eqnarray}
which means the sign of $v_n$, i.e. the direction of the mean velocity is determined by the factor $\Gamma$ if $U>0$ is assumed.

\bibliography{poly2,colloid,stokes,neu,kanso,andre}

\end{document}